\documentclass[prd,11pt,superscriptaddress,nofootinbib,twoside,raggedbottom]{revtex4-1}
\pdfoutput=1

\usepackage[utf8]{inputenc}

\usepackage{footmisc}

\usepackage[colorlinks]{hyperref}
\hypersetup{
    colorlinks   = true,
    citecolor    = red,
	linkcolor    = blue
}

\usepackage{bbm}
\usepackage{amsfonts}
\usepackage{mathrsfs}

\usepackage{bbold}
\usepackage{amsmath,amssymb}

\usepackage{graphicx}               
\usepackage{url}
\usepackage{float}
\usepackage{color}

\newcommand{\be}{\begin{equation}}
\newcommand{\ee}{\end{equation}}
\newcommand{\bea}{\begin{eqnarray}}
\newcommand{\eea}{\end{eqnarray}}
\newcommand{\beq}{\begin{eqnarray}}
\newcommand{\eeq}{\end{eqnarray}}

\newcommand{\MM}{\mathcal{M}}
\newcommand{\OO}{\mathcal{O}}

\def\ve{v_\text{e}}
\def\vve{\vect{v}_\text{e}}
\def\vesc{v_\text{esc}}
\def\vect#1{\boldsymbol{#1}}

\begin{document}

\title{Multi-Channel Direct Detection of Light Dark Matter: Theoretical Framework}

%
%
%

%
\author{Tanner~Trickle}
\author{Zhengkang~Zhang}
\affiliation{Department of Physics, University of California, Berkeley, CA 94720, USA}
\affiliation{Theoretical Physics Group, Lawrence Berkeley National Laboratory, Berkeley, CA 94720, USA}

\author{Kathryn~M.~Zurek}
\affiliation{Walter Burke Institute for Theoretical Physics, California Institute of Technology, Pasadena, CA 91125, USA}

\author{Katherine~Inzani}
\author{Sin\'ead~M.~Griffin}
\affiliation{Materials Sciences Division, Lawrence Berkeley National Laboratory, Berkeley, CA 94720, USA}
\affiliation{Molecular Foundry, Lawrence Berkeley National Laboratory, Berkeley, CA 94720, USA}

\begin{abstract}
We present a unified theoretical framework for computing spin-independent direct detection rates via various channels relevant for sub-GeV dark matter -- nuclear recoils, electron transitions and single phonon excitations. Despite the very different physics involved, in each case the rate factorizes into the particle-level matrix element squared, and an integral over a target material- and channel-specific dynamic structure factor. We show how the dynamic structure factor can be derived in all three cases following the same procedure, and extend previous results in the literature in several aspects. For electron transitions, we incorporate directional dependence and point out anisotropic target materials with strong daily modulation in the scattering rate. For single phonon excitations, we present a new derivation of the rate formula from first principles for generic spin-independent couplings, and include the first calculation of phonon excitation through electron couplings. We also discuss the interplay between single phonon excitations and nuclear recoils, and clarify the role of Umklapp processes, which can dominate the single phonon production rate for dark matter heavier than an MeV. Our results highlight the complementarity between various search channels in probing different kinematic regimes of dark matter scattering, and provide a common reference to connect dark matter theories with ongoing and future direct detection experiments.
\end{abstract}

\maketitle

\tableofcontents
\thispagestyle{empty}
\newpage

\section{Introduction}
\label{sec:intro}

Direct detection has been playing a central role in the quest for the particle nature of dark matter (DM). Over the past few decades, tremendous progress has been made at a range of experiments focused on nuclear recoil signals, including ANAIS \cite{Amare:2019jul}, CRESST \cite{Cozzini2002,Petricca:2017zdp,Angloher:2015ewa}, DAMA/LIBRA \cite{Baum:2018ekm}, DAMIC \cite{deMelloNeto:2015mca,AguilarArevalo2019a}, DarkSide-50 \cite{Agnes2018b}, DM-Ice \cite{Jo:2016qql}, KIMS \cite{Kim:2015prm}, LUX~\cite{Akerib:2018kjf,Akerib2019a,Akerib2019b}, SABRE \cite{Shields:2015wka}, SuperCDMS \cite{Agnese:2014aze,Agnese:2016cpb,Agnese:2015nto,Agnese2018a,Agnese2019a,Agnese2018b}, and XENON1T~\cite{Aprile2018a, Aprile:2019xxb}. While these experiments have excluded much of the parameter space for DM heavier than roughly a GeV, much less is known about lighter DM. For sub-GeV DM, conventional nuclear recoil searches lose sensitivity due to kinematic mismatch, as only a small fraction of DM's kinetic energy can be deposited on the heavier nuclei. Even with next generation detectors sensitive to sub-eV energy depositions, nuclear recoils can at best probe DM masses down to $\OO(100\,\text{MeV})$. 

To cover a broader mass range, electrons have been considered as an alternate pathway to detecting light DM. A variety of targets have been studied, including noble gas atoms which can be ionized with $\OO(10\,\text{eV})$ energy deposition, semiconductors where electron transitions can happen across $\OO(\text{eV})$ band gaps~\cite{Essig:2011nj,Graham:2012su,Essig:2012yx,Lee:2015qva,Essig:2015cda,Derenzo:2016fse,Hochberg:2016sqx,Bloch:2016sjj,Essig:2017kqs,Kurinsky:2019pgb}, as well as systems with $\OO(\text{meV})$ gaps like superconductors~\cite{Hochberg:2015pha,Hochberg:2015fth,Hochberg:2016ajh} and Dirac materials~\cite{Hochberg:2017wce,Coskuner:2019odd,Geilhufe:2019ndy}. Electron transitions can potentially extract all of DM's kinetic energy, and thus constitute a more efficient search channel than nuclear recoils. For example, semiconductor targets can probe DM masses down to $\OO(\text{MeV})$.

When the energy deposition is below the band gap, electron transitions are kinematically forbidden. However, there are condensed matter systems with collective excitations that can couple to the DM. For example, collective excitations in superfluid helium (phonons and rotons) are sensitive to $\OO(\text{meV})$ energy depositions, especially via phonon pair production~\cite{Schutz:2016tid,Knapen:2016cue,Acanfora:2019con,Caputo:2019cyg}. In a crystal target, the active degrees of freedom below the electronic band gap are acoustic and optical phonons -- quanta of collective oscillations of atoms/ions. Direct excitation of single phonons in crystals has been recently proposed as a new search channel for light DM~\cite{Knapen:2017ekk,Griffin:2018bjn}. Optical phonons typically have energies of $\OO(10\text{-}100\,\text{meV})$, and can be excited by DM as light as $\OO(10\,\text{keV})$. Acoustic phonons are gapless and, assuming an $\OO(\text{meV})$ detector threshold, can also probe DM down to $\OO(10\,\text{keV})$.

\begin{figure}[ht!]
\includegraphics[width=0.6\linewidth]{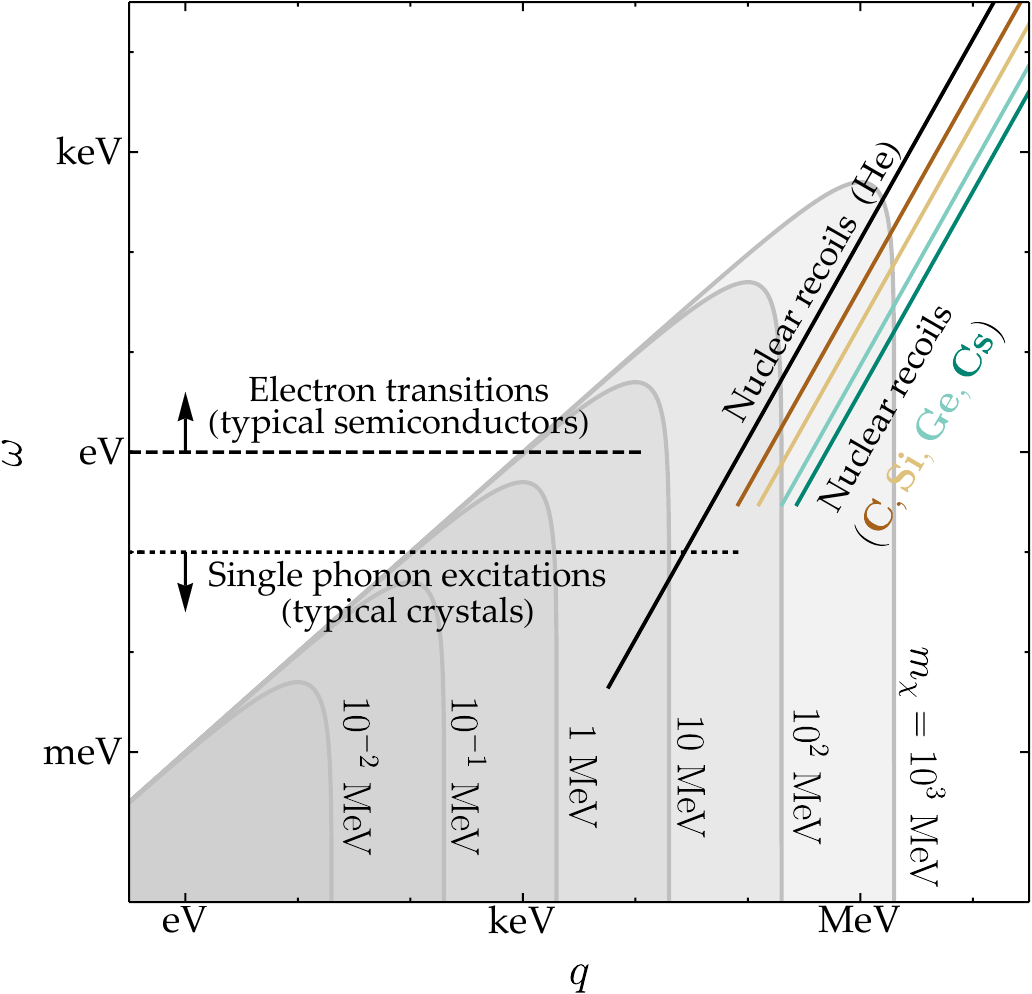}
\caption{\label{fig:kin}
Illustration of kinematic regimes probed via the three detection channels considered in this paper. For an incoming DM particle with velocity $v=10^{-3}$, the momentum transfer $q$ and energy deposition $\omega$ are bounded by $\omega\le qv-q^2/2m_\chi$, shown by the shaded regions for three DM masses. Nuclear recoils require $\omega=q^2/2m_N$ for a given type of nucleus, shown by the solid lines for helium and several elements in existing or proposed crystal targets. Standard calculations assuming scattering off individual nuclei break down below a few meV (a few hundred meV) for superfluid He (crystal targets), where we truncate the lines. Electron transitions can be triggered for $\omega$ above the band gap, which is $\OO(\text{eV})$ for typical semiconductors, as shown by the dashed line. The end point at $q\sim 10\,$keV corresponds to a few times $\alpha m_e$, above which valence electron wavefunctions are suppressed, and only (semi-)core electrons can contribute (which requires $\omega$ to be much higher than the band gap). Single phonon excitations are relevant for $\omega\lesssim\OO(100\,\text{meV})$ in typical crystals, as shown by the dotted line. The momentum transfer can be up to $q\sim \sqrt{m_N\omega_\text{ph}}\sim\OO(100\,\text{keV})$ with $\omega_\text{ph}$ the phonon energies, above which the rate is suppressed by the Debye-Waller factor. We see that a GeV-mass DM can be probed by all three channels; a 10\,MeV DM is out of reach in conventional nuclear recoil searches, but can be searched for via electron transitions in semiconductors and single phonon excitations in crystals; a sub-MeV DM cannot even trigger electron transitions in eV-gap materials, but can still be detected via single phonon excitations.
}
\end{figure}

All these detection channels do not exist in isolation. Depending on the DM mass and couplings to Standard Model (SM) particles, it may either cause nuclear recoils, or induce electron transitions, or excite phonons in the same target material. Thus, when designing direct detection experiments, an important consideration should be to search for DM across multiple channels in parallel. The kinematic interplay between several channels that we will discuss in detail is illustrated in Fig.~\ref{fig:kin}.

\clearpage

On the theory side, most of the basic ingredients for the rate calculation are known. However, they have been developed in separate contexts, and at first sight look very different for different detection channels. In our opinion, it would be much more convenient to have a common theoretical framework for all these calculations. This will not only facilitate the comparison of target materials across various existing and proposed search channels, but also provide the necessary calculation tools when new search channels are considered in the future.

It is the purpose of this paper to lay out such a formalism, focusing on spin-independent (SI) DM interactions.\footnote{The idea of treating various detection channels in a common framework was previously advocated in Ref.~\cite{Coskuner:2018are}, where the focus was on DM nuggets. Here we follow the same spirit and develop a formalism for calculating direct detection rates for general DM models, assuming a point-like DM particle.} As we will see, for each detection channel, the calculation is factorized into a particle physics model-specific part and a target response-specific part. The latter is encoded in a dynamic structure factor, to be computed by quantizing the particle number density operators in the Hilbert space of the excitations under study. We show how this is done in three cases -- nuclear recoils, electron transitions and single phonon excitations. While the first two are relatively simple, and our calculation is mostly a formal rederivation of known results, the phonon calculation presented here contains new aspects. Our general framework allows us to derive single phonon excitation rates for arbitrary SI couplings from first principles, such as phonon excitation by coupling to electrons.

In addition to deriving general rate formulae in this unified framework, we also aim to clarify various conceptual and technical issues in direct detection calculations, and present new results that highlight some previously overlooked experimental prospects. For nuclear recoils, we clarify the range of validity of the standard calculation. For electron transitions, we go beyond the commonly made isotropic approximation. In fact, there exist simple materials with large anisotropies. As an example, we consider boron nitride (BN) with a hexagonal crystal structure, and $\OO(\text{eV})$ band gap, and show that the expected rate can vary by $\pm (10\,\text{-}\,40)\,\%$ during a day as the DM wind enters from different directions. Such daily modulation signals have been pointed out previously for electron transitions in graphene~\cite{Hochberg:2016ntt}, carbon nanotubes~\cite{Cavoto:2017otc} and Dirac materials such as ZrTe$_5$ and BNQ-TTF~\cite{Coskuner:2019odd,Geilhufe:2019ndy}, and for single phonon excitations in sapphire~\cite{Griffin:2018bjn} where they help distinguish signal from background. Here we show that also $\OO(\text{eV})$ band gap three dimensional semiconductors, like BN, can exhibit daily modulation.\footnote{See also Refs.~\cite{Kadribasic:2017obi,Budnik:2017sbu,Rajendran:2017ynw} for proposals that take advantage of direction-dependent threshold effects.} Finally, for single phonon excitations, we extend the rate calculation to DM heavier than an MeV, where the DM's de Broglie wavelength is shorter than the typical lattice spacing, and Umklapp processes can contribute significantly. We point out an interesting interplay with nuclear recoils, and demonstrate the complementarity between the two channels. We also compute the phonon production rate for generic couplings to the proton, neutron and electron, extending previous results for dark photon mediated interactions.

We focus on the theoretical framework in the present work; in a companion paper~\cite{Griffin:2019mvc}, we apply the results presented here to carry out a comparative study of many candidate target materials, and discuss strategies to optimize the search across multiple channels. We also note that there are additional detection channels beyond those we discuss in detail here ({\em e.g.}\ excitation of molecular states~\cite{Essig:2016crl,Arvanitaki:2017nhi,Essig:2019kfe}, multi-excitation production in superfluid helium~\cite{Schutz:2016tid,Knapen:2016cue,Acanfora:2019con,Caputo:2019cyg}), which have been pursued and can be studied in the same framework.

\section{General Framework for Spin-Independent Dark Matter Scattering}
\label{sec:general}

In a direct detection event, a non-relativistic DM particle, $\chi$, deposits a certain amount of energy, and triggers a transition $|i\rangle\to|f\rangle$ in the target system. We assume the target system is initially prepared in an energy eigenstate $|i\rangle$ (usually the ground state) and, as usual, treat the incoming and outgoing DM particles as momentum eigenstates $|\vect{p}\rangle$, $|\vect{p}'\rangle$, with $\vect{p}=m_\chi\vect{v}$, $\vect{p}'=\vect{p}-\vect{q}$. For a given incoming velocity $\vect{v}$ and momentum transfer (from the DM to the target) $\vect{q}$, the energy deposition is
\be
\omega_{\vect{q}} = \frac{1}{2}m_\chi v^2 -\frac{(m_\chi\vect{v}-\vect{q})^2}{2m_\chi} = \vect{q}\cdot\vect{v} -\frac{q^2}{2m_\chi}\,.
\ee
Here and in what follows, we denote $q\equiv|\vect{q}|$, where $\vect{q}$ is the momentum 3-vector. Note that for given DM mass $m_\chi$, the energy deposition is bounded by the parabola, $\omega_{\vect{q}}\le qv_\text{max}-q^2/2m_\chi$, as shown in Fig.~\ref{fig:kin}. Applying Fermi's Golden Rule and summing over the final states, we obtain the rate:
\be
\Gamma (\vect{v}) = V \int\frac{d^3q}{(2\pi)^3} \sum_f\, \bigl| \langle \vect{p}', f | \,\delta \hat H \, | \vect{p} , i \rangle \bigr|^2\, 2 \pi \delta\bigl(E_f-E_i-\omega_{\vect{q}}\bigr)\,,
\ee
where $\delta \hat H$ is the interaction Hamiltonian, $| \vect{p} , i \rangle = | \vect{p} \rangle \otimes | i \rangle$, $| \vect{p}' , f \rangle = | \vect{p}' \rangle \otimes | f \rangle$. We take the quantum states to be unit normalized unless specified otherwise, {\em e.g.}\ $\langle \vect{p} | \vect{p} \rangle = \langle i | i \rangle = 1$.

The DM part of the matrix element can be evaluated universally at the Born level:
\be
\langle \vect{p}'| \,\delta \hat H \,| \vect{p}\rangle = \frac{1}{V}\int d^3x \,e^{i\vect{q}\cdot\vect{x}} \,{\cal V}(\vect{x}) = \frac{1}{V}\,{\cal \widetilde V}(-\vect{q})\,,
\ee
where $V$ is the total spatial volume, ${\cal V}(\vect{x})$ is the effective scattering potential felt by the DM, and ${\cal \widetilde V}$ is its Fourier transform. We focus on SI couplings in the present work, in which case the scattering potential takes the form\footnote{More generally, DM interactions can be classified by nonrelativistic effective operators~\cite{Chang:2009yt,Fitzpatrick:2010br,Fitzpatrick:2012ix,Gresham:2014vja}. The SI interaction we focus on here is the leading operator if generated without velocity suppression. Other operators result in spin and/or velocity dependence of the scattering potential ${\cal V}(\vect{x})$, and may be probed via additional detection channels beyond those considered in this work. For example, DM coupling to the electron spin can excite magnons in solid state systems with magnetic order~\cite{Trickle:2019ovy}. We leave a general effective field theory study of light DM direct detection to future work.}
\be
{\cal V}(\vect{x}) = \int d^3x' \,\bigl[ n_p(\vect{x}') {\cal V}_p(\vect{x} - \vect{x}') + n_n(\vect{x}') {\cal V}_n(\vect{x} - \vect{x}') + n_e(\vect{x}') {\cal V}_e(\vect{x}-\vect{x}') \bigr] .
\label{eq:vx}
\ee
Here $n_p, n_n, n_e$ are the proton, neutron and electron number densities in the target, and $\mathcal{V}_p, \mathcal{V}_n, \mathcal{V}_e$ are the respective scattering potentials from a single particle located at the origin. We thus have
\be
{\cal \widetilde V}(-\vect{q}) = \widetilde n_p(-\vect{q}) \,\widetilde{\mathcal{V}}_p(q) + \widetilde n_n(-\vect{q}) \,\widetilde{\mathcal{V}}_n(q) + \widetilde n_e(-\vect{q}) \,\widetilde{\mathcal{V}}_e(q) \,.
\label{eq:Vq}
\ee
Note that for SI interactions, $\widetilde{\cal V}_\psi(-\vect{q})=\widetilde{\cal V}_\psi(q)$ ($\psi=p,n,e$) are functions of only the magnitude of $\vect{q}$. In vacuum, they simply coincide with $2\to 2$ scattering matrix elements $\mathcal{M}_{\chi\psi}(q)$ familiar from standard quantum field theory calculations. In the target medium, however, they may receive corrections due to screening effects (see Sec.~\ref{sec:general-medium}). We can define (momentum-dependent) effective in-medium couplings $f_p, f_n, f_e$ to account for screening effects, while the corresponding couplings in the vacuum Lagrangian are denoted by $f_p^0, f_n^0, f_e^0$. We can write
\be
\widetilde{\mathcal{V}}_\psi(-\vect{q}) = \frac{f_\psi(\vect{q})}{f_\psi^0} \mathcal{M}_{\chi\psi}(q) \equiv f_\psi(\vect{q})\, \mathcal{M}_0(q)\,,
\ee
where $\MM_0 = \MM_{\chi p}/f_p^0 = \MM_{\chi n}/f_n^0 = \MM_{\chi e}/f_e^0$ is the vacuum matrix element for DM scattering off any of the constituent particles (proton, neutron or electron) with unit coupling.  The total scattering potential is then
\be
{\cal \widetilde V}(-\vect{q}) = \left[ f_p(\vect{q})\, \widetilde n_p(-\vect{q}) + f_n(\vect{q})\, \widetilde n_n(-\vect{q}) + f_e(\vect{q})\, \widetilde n_e(-\vect{q}) \right] \mathcal{M}_0(q) \,.
\label{eq:Vq2}
\ee
Let us rewrite this equation as follows:
\beq
{\cal \widetilde V}(-\vect{q}) &=& \mathcal{M}_{\chi n}(q) \biggl[ \frac{f_p(\vect{q})\, \widetilde n_p(-\vect{q}) + f_n(\vect{q})\, \widetilde n_n(-\vect{q}) + f_e(\vect{q})\, \widetilde n_e(-\vect{q})}{f_n^0} \biggr] \\
&=& \mathcal{M}_{\chi e}(q) \biggl[ \frac{f_p(\vect{q})\, \widetilde n_p(-\vect{q}) + f_n(\vect{q})\, \widetilde n_n(-\vect{q}) + f_e(\vect{q})\, \widetilde n_e(-\vect{q})}{f_e^0} \biggr]\,.
\eeq
Depending on the DM model and the process under consideration, we will factor out either $\mathcal{M}_{\chi n}$ or $\mathcal{M}_{\chi e}$, and define a target form factor, ${\cal F}_T(\vect{q})$, composed of contributions from protons, neutrons and electrons, as the quantity in brackets. In other words, we have
\be
{\cal \widetilde V}(-\vect{q}) = \MM(q)\, {\cal F}_T(\vect{q})\,,
\label{eq:factorization}
\ee
where $\MM$ stands for $\MM_{\chi n}$ or $\MM_{\chi e}$. We can further factor out the $q$ dependence of $\MM$, which can only come from the mediator propagator for tree-level scattering:
\beq
\MM(q) &=& \MM(q_0)\, {\cal F}_\text{med}(q)\,,\\
{\cal F}_\text{med}(q) &=& 
\begin{cases}
1 & \text{(heavy mediator)},\\
(q_0/q)^2 & \text{(light mediator)}.
\end{cases}
\label{eq:Fmed}
\eeq
The reference momentum transfer is conventionally chosen to be $q_0=m_\chi v_0$ (with $v_0$ the DM's velocity dispersion) for DM-neutron scattering, and $q_0=\alpha m_e$ for DM-electron scattering.

The factorization in Eq.~\eqref{eq:factorization} is a key component of the formalism. From the target-independent particle-level matrix element $\MM$, we define the reference cross sections:
\be
\overline\sigma_n \equiv \frac{\mu^2_{\chi n}}{\pi} \overline{|\MM_{\chi n}(q_0)|^2}_{q_0=m_\chi v_0}\,,
\qquad
\overline\sigma_e \equiv \frac{\mu^2_{\chi e}}{\pi} \overline{|\MM_{\chi e}(q_0)|^2}_{q_0=\alpha m_e}\,,
\ee
where $\mu$ denotes the reduced mass. These coincide with the total cross sections of DM-neutron and DM-electron scattering in the heavy mediator case. On the other hand, ${\cal F}_T$ is target specific, from which we define the {\it dynamic structure factor}:\footnote{Here we adopt a slightly different normalization convention compared to Ref.~\cite{Coskuner:2018are}. The right hand side of Eq.~\eqref{eq:sfun} here is identified with $\frac{2\pi}{\Omega}S(\vect{q},\omega)$ in Ref.~\cite{Coskuner:2018are}, where $\Omega$ is the primitive cell volume.}
\be
S(\vect{q},\omega) \equiv \frac{1}{V} \sum_f \bigl|\langle f| {\cal F}_T(\vect{q}) |i\rangle \bigr|^2 \,2\pi \delta\bigl(E_f-E_i-\omega\bigr),\label{eq:sfun}
\ee
which encapsulates response of the target to DM couplings to the proton, neutron and electron. Combining the two parts, we have
\be
\Gamma (\vect{v}) = \frac{\pi\overline\sigma}{\mu^2} \int\frac{d^3q}{(2\pi)^3} \,{\cal F}_\text{med}^2(q)\, S\bigl(\vect{q},\omega_{\vect{q}}\bigr)\,,
\label{eq:gamma}
\ee
where $\bar\sigma, \mu$, again, denote either $\bar\sigma_n, \mu_{\chi n}$ or $\bar\sigma_e, \mu_{\chi e}$.

Let us highlight the following regarding the dynamic structure factor $S(\vect{q},\omega)$.
\begin{itemize}
\item $S(\vect{q},\omega)$ captures the target's response to an energy-momentum deposition $(\vect{q},\omega)$. 
\item $S(\vect{q},\omega)$ depends on the distribution of constituent particles $p,n,e$ in the target system via $\widetilde n_p, \widetilde n_n, \widetilde n_e$, which in turn depends on the nucleus types and electron wavefunctions. It is therefore {\it target material specific}.
\item $S(\vect{q},\omega)$ also depends on the active degrees of freedom in the target system via the choice of $|f\rangle$, which in turn determines how ${\cal F}_T(\vect{q})$ should be quantized. It is therefore {\it excitation (detection channel) specific}.
\item If only one of the constituent particles $p, n, e$ is responsible for the transitions $|i\rangle\to|f\rangle$, $S(\vect{q},\omega)$ is DM model independent. Otherwise it depends on ratios (but not the overall strength) of the couplings $f_p^0, f_n^0, f_e^0$.
\item For any given DM mass $m_\chi$ and incoming velocity $\vect{v}$, only a slice in the $(\vect{q},\omega)$ space, $\omega=\omega_{\vect{q}}$, is probed in the scattering process. The parabolic boundary of kinematic region for each $m_\chi$ in Fig.~\ref{fig:kin} is the envelope of these slices for all $\vect{v}$ directions for fixed magnitude of $\vect{v}$.
\end{itemize}

Finally, to obtain the total rate per target mass, we average over the DM's initial velocity, multiply by the number of DM particles in the detector, and divide by the detector mass, giving
\be
R = \frac{1}{\rho_T} \frac{\rho_\chi}{m_\chi} \int d^3v\, f_\chi(\vect{v})\,\Gamma(\vect{v})\,,
\ee
where $\rho_T$ is the target mass density, $\rho_\chi$ is the local DM energy density, and $f_\chi$ is the DM's velocity distribution in the target rest frame. A common choice for $f_\chi$ is a truncated Maxwell-Boltzmann (MB) distribution boosted by the Earth's velocity with respect to the galactic rest frame,
\beq
f_\chi^\text{MB}(\vect{v}) &=& \frac{1}{N_0} e^{-(\vect{v}+\vve)^2/v_0^2} \,\Theta \bigl(\vesc-|\vect{v} +\vve|\bigr) \,,\label{eq:fmb}\\
N_0 &=& \pi^{3/2} v_0^2 \Biggl[ v_0\,\text{erf} \bigl(\vesc/v_0\bigr) -\frac{2\,\vesc}{\sqrt{\pi}} \exp\bigl(-\vesc^2/v_0^2\bigr)\Biggr] .
\eeq
In the calculations presented in this paper, we take $\rho_\chi=0.4\,\text{GeV}/\text{cm}^3$, $v_0=230\,\text{km}/\text{s}$, $\vesc=600\,\text{km}/\text{s}$, $\ve=240\,\text{km}/\text{s}$.

In addition to the total rate, it is often useful to know the differential rate with respect to the energy deposition onto the target $\omega$. This simply requires inserting delta functions into the integrals to pick out the contributions with $\omega=\omega_{\vect{q}}$:
\beq
\frac{d\Gamma}{d\omega} &=& \frac{\pi\overline\sigma}{\mu^2} \int\frac{d^3q}{(2\pi)^3} \,{\cal F}_\text{med}^2(q)\, S\bigl(\vect{q},\omega_{\vect{q}}\bigr)\,\delta\bigl(\omega-\omega_{\vect{q}}\bigr)\,,\\
\frac{dR}{d\omega} &=& \frac{1}{\rho_T} \frac{\rho_\chi}{m_\chi} \int d^3v \, f_\chi(\vect{v})\,\frac{d\Gamma}{d\omega}\,.
\label{eq:drded}
\eeq

To summarize, we have the following algorithm for computing the rate for a given detection channel.
\begin{itemize}
\item First, identify the initial and final states $|i\rangle, |f\rangle$ according to the type of excitation.
\item Next, quantize ${\cal F}_T(\vect{q})$ in terms of the relevant degrees of freedom such that it acts on the target Hilbert space to induce the transitions $|i\rangle \to |f\rangle$.
\item Then, compute the transition matrix element $\langle f| {\cal F}_T(\vect{q}) |i\rangle$, and thus the dynamic structure factor $S(\vect{q},\omega)$ via Eq.~\eqref{eq:sfun}.
\item Finally, obtain the (differential) rate via Eqs.~\eqref{eq:gamma}-\eqref{eq:drded}.
\end{itemize}
We will carry out this procedure for each detection channel in the next three sections. Before doing so, let us discuss some technical details regarding the phase space integration and in-medium effects.

\subsection{Phase Space Integration}
\label{sec:general-integration}

We see from Eqs.~\eqref{eq:gamma}-\eqref{eq:drded} that once the dynamic structure factor $S(\vect{q},\omega)$ is known, we need to perform a six-dimensional integral over $\vect{v}$ and $\vect{q}$ to obtain the event rate $R$. The integration gives familiar results in the special case of isotropic target response, but is more complicated in the general anisotropic case. We now discuss the two cases in turn.

\paragraph*{a) Special case: isotropic target response.}

If $\bigl|\langle f| {\cal F}_T(\vect{q}) |i\rangle \bigr|^2=\bigl|\langle f| {\cal F}_T(q) |i\rangle \bigr|^2$, as is the case for nuclear recoils, the only dependence on the direction of $\vect{q}$ is from the $\delta$-function,
\be
\delta\bigl(E_f-E_i-\omega_{\vect{q}}\bigr)=\frac{1}{qv}\delta\biggl(\cos\theta_{qv}-\frac{q}{2m_\chi v}-\frac{E_f-E_i}{qv}\biggr)\,,
\ee
where $\theta_{qv}$ is the angle between $\vect{q}$ and $\vect{v}$. Integrating over the angular variables, we have
\be
\Gamma (\vect{v}) = \frac{\overline\sigma}{2\mu^2 v} \int qdq \,{\cal F}_\text{med}^2(q)\, \frac{1}{V} \sum_f \bigl|\langle f| {\cal F}_T(q) |i\rangle \bigr|^2 \, \Theta\bigl(v - v_\text{min}(q,E_f-E_i)\bigr)\,,\label{eq:Gamma-iso}
\ee
where
\be
v_\text{min} (q,\omega) = \frac{q}{2m_\chi} +\frac{\omega}{q}\,.
\ee
The velocity integral then gives
\be
R = \frac{1}{\rho_T} \frac{\rho_\chi}{m_\chi} \frac{\overline\sigma}{2\mu^2} \int qdq \,{\cal F}_\text{med}^2(q)\, \frac{1}{V} \sum_f \bigl|\langle f| {\cal F}_T(q) |i\rangle \bigr|^2 \,\eta\bigl(v_\text{min}(q,E_f-E_i)\bigr)\,,\label{eq:R-iso}
\ee
where
\be
\eta(v_\text{min}) = \int d^3v \, \frac{f_\chi(\vect{v})}{v} \,\Theta(v - v_\text{min})\,.\label{eq:etafun}
\ee

These results are familiar from the standard nuclear recoil calculation~\cite{Lin:2019uvt}, and have also been used in previous electron transition calculations, where the target response has been assumed to be isotropic. Note that they hold for any DM velocity distribution $f_\chi(\vect{v})$. In the case of the MB distribution in Eq.~\eqref{eq:fmb}, the $\eta$ function can be evaluated analytically, giving
\be
\eta_\text{MB}(v_\text{min}) = 
\begin{cases}
\frac{\pi v_0^2}{2N_0} \Bigl\{ \sqrt{\pi}\, \frac{v_0}{\ve} \Bigl[ \text{erf}\bigl(\frac{v_\text{min}+\ve}{v_0}\bigr) -\text{erf}\bigl(\frac{v_\text{min}-\ve}{v_0}\bigr) \Bigr] -4 \exp\bigl(-\frac{\vesc^2}{v_0^2}\bigr) \Bigr\} \\[4pt] 
\qquad\qquad\qquad\qquad\qquad\qquad\qquad\qquad \text{if}\quad v_\text{min} < \vesc-\ve\,, \\[12pt]
\frac{\pi v_0^2}{2N_0} \Bigl\{ \sqrt{\pi}\, \frac{v_0}{\ve} \Bigl[ \text{erf}\bigl(\frac{\vesc}{v_0}\bigr) -\text{erf}\bigl(\frac{v_\text{min}-\ve}{v_0}\bigr) \Bigr] -2\,\bigl(\frac{\vesc-v_\text{min}+\ve}{\ve}\bigr) \exp\bigl(-\frac{\vesc^2}{v_0^2}\bigr) \Bigr\} \\[4pt]
\qquad\qquad\qquad\qquad\qquad\qquad\qquad\qquad \text{if}\quad \vesc-\ve < v_\text{min} < \vesc-\ve\,, \\[12pt]
0  \qquad\qquad\qquad\qquad\qquad\qquad\qquad\quad\;\; \text{if}\quad v_\text{min} > \vesc+\ve\,.
\end{cases}
\label{eq:etamb}
\ee
We see that five of the six integrals have been done analytically, and we are left only with a one-dimensional integral over $q$ (which can also be done analytically in the case of nuclear recoils).

\paragraph*{b) General case: anisotropic target response.}

Generally, crystal targets are not fully isotropic, as the crystal structures break rotation symmetries. This implies that, for a terrestrial detector, since the DM wind comes in from different directions at different times of the day, there can be daily modulation in the detection rate. While the existence of this effect is well-known~\cite{Essig:2011nj,Essig:2015cda,Knapen:2017ekk}, it has been calculated only recently in the contexts of single phonon excitations~\cite{Griffin:2018bjn} and electron transitions in Dirac materials~\cite{Coskuner:2019odd,Geilhufe:2019ndy}, where the energy deposition is $\OO(\text{meV})$. In Sec.~\ref{sec:electron-modulation}, we calculate this effect for the first time in electron transitions in an $\OO(\text{eV})$ gap target. 

When $\bigl|\langle f| {\cal F}_T(\vect{q}) |i\rangle \bigr|^2$ depends on the direction of $\vect{q}$, the six-dimensional integral generally does not admit a simple analytical solution. To proceed, we first evaluate the velocity integral and define~\cite{Griffin:2018bjn,Knapen:pc}
\be
g(\vect{q},\omega) \equiv \int d^3 v \, f_\chi(\vect{v}) \,2\pi \delta(\omega-\omega_{\vect{q}}) \,.
\label{eq:gfun}
\ee
The rate can then be written in terms of this $g(\vect{q}, \omega)$ function as
\be
R = \frac{1}{\rho_T} \frac{\rho_\chi}{m_\chi} \frac{\pi\overline\sigma}{\mu^2} \int\frac{d^3q}{(2\pi)^3} \,{\cal F}_\text{med}^2(q)\,\frac{1}{V} \sum_f \bigl|\langle f| {\cal F}_T(\vect{q}) |i\rangle \bigr|^2 \,g(\vect{q},E_f-E_i) \,.
\ee

For general velocity distributions $f_\chi$, we still have to evaluate a six-dimensional integral, which is a numerically intensive task. However, for the commonly assumed MB distribution, Eq.~\eqref{eq:fmb}, the $g(\vect{q}, \omega)$ function can be evaluated analytically, giving
\be
g(\vect{q},\omega) = \frac{2\pi^2 v_0^2}{N_0 q} \Bigl[\exp \bigl(-v_-^2/v_0^2\bigr) -\exp \bigl(-\vesc^2/v_0^2\bigr)\Bigr] \,,
\ee
where
\be
v_- = \min \biggl\{ \,\frac{1}{q} \,\biggl|\vect{q}\cdot\vect{v}_\text{e}+\frac{q^2}{2m_\chi}+\omega\biggr|\,,\, \vesc \biggr\}\,.
\ee
Thus, only the three-dimensional integral over $\vect{q}$ needs to be done numerically (in addition to other integrals that may be encountered in the evaluation of the dynamic structure factor).

\subsection{In-Medium Effects}
\label{sec:general-medium}

In the case of a vector or scalar mediator, in-medium effects can cause screening and affect direct detection rates. They must be taken into account when deriving the target response ${\cal F}_T(\vect{q})$ (and hence the dynamical structure factor $S(\vect{q},\omega)$) when present. While the treatment of in-medium effects has been discussed in various contexts~\cite{Hochberg:2015fth,Hochberg:2017wce,Knapen:2017xzo,Coskuner:2019odd}, we review it here for completeness. In particular, we derive the screening factors $f_\psi(\vect{q})/f_\psi^0$ ($\psi = p, n, e$) in this subsection.

For nonrelativistic systems relevant for direct detection that we focus on here, only electrons can contribute significantly to screening when the energy deposition is above phonon frequencies ($\omega \gtrsim \mathcal{O}(100 \text{ meV})$, corresponding to $m_\chi \gtrsim \mathcal{O}(100 \text{ keV})$), as nuclei are too heavy to respond. At lower frequencies that match energy depositions in phonon excitation processes, there is additional screening in an ionic (polar) crystal due to relative motion of ions. However, as we will see in Sec.~\ref{sec:phonon-e}, the ions' response should be included in the source term in Maxwell's equations in order to be quantized in terms of phonon modes. Thus, also in this case, we consider only electron contributions to in-medium effects.\footnote{In-medium effects are also important when deriving astrophysical and cosmological constraints on vector mediators~\cite{An:2013yfc,Hardy:2016kme,Knapen:2017xzo}, where other SM particles may be relevant.}

Consider a vector mediator $A'$, and suppose the vacuum Lagrangian takes the form
\beq
\mathcal{L} &=& -\frac{1}{4} F_{\mu \nu} F^{\mu \nu} + e J^\mu_p A_\mu - e J^\mu_e A_\mu \nonumber \\
&& -\frac{1}{4} F^{'}_{\mu \nu} F^{'\mu \nu} + \frac{1}{2}m_{A'}^2 A'_\mu A'^\mu + g_\chi J_\chi^\mu A'_\mu \nonumber \\
&& +\bigl( f_p^0 J_p^\mu + f_n^0 J_n^\mu + f_e^0 J^\mu_e \bigr) A'_\mu \, ,
\label{eq:Lvac}
\eeq
where $J_\psi^\mu = \bar \psi \gamma^\mu \psi$ ($\psi=p,n,e$). Here the first line is standard electromagnetism, the second line is the dark sector Lagrangian, and the third line contains $A'$ couplings to SM particles. We assume $|f_\psi^0| \ll 1$, and consistently keep terms only at linear order in these couplings. Because the electron current $J_e^\mu$ couples to the linear combination $A_\mu+\kappa A'_\mu$, with $\kappa = -f_e^0/e$, as opposed to just $A_\mu$, the in-medium photon self-energy $\Pi^{\mu\nu}(\vect{q})$ implies the following terms in the momentum space quantum effective action,
\be
\frac{1}{2}\,\Pi^{\mu\nu} (A_\mu +\kappa A'_\mu)(A_\nu +\kappa A'_\nu) = \frac{1}{2}\,\Pi^{\mu\nu} A_\mu A_\nu +\kappa\,\Pi^{\mu\nu} A_\mu A'_\nu +\OO(\kappa^2) \,.
\ee
As in Ref.~\cite{Coskuner:2019odd}, we can project $\Pi^{\mu\nu}$ onto the three polarizations, 
\be
\epsilon_L^\mu = \frac{1}{\sqrt{q^\alpha q_\alpha}} \bigl(q,\, \omega\vect{\hat q}\bigr) \,,\qquad
\epsilon_\pm^\mu = \frac{1}{\sqrt{2}} \bigl(0,\, \vect{\hat e}_\perp \pm i(\vect{\hat q}\times\vect{\hat e}_\perp) \bigr)\,,
\ee
(where $\vect{\hat q}=\vect{q}/|\vect{q}|$, and $\vect{\hat e}_\perp$ is a unit vector perpendicular to $\vect{q}$), and diagonalize the $3\times3$ matrix
\be
{\cal K}_{\lambda\lambda'} \equiv -\epsilon_\lambda^{\mu*}\Pi_{\mu\nu}\epsilon_{\lambda'}^\nu\,,
\ee
to find the canonical modes. It is worth noting that the polarization vectors satisfy
\be
g_{\mu\nu}\epsilon_\lambda^{*\mu}\epsilon_{\lambda'}^\nu = -\delta_{\lambda\lambda'}\,,\qquad
\sum_{\lambda} \epsilon_\lambda^\mu \epsilon_\lambda^{\nu*} = -\biggl(g^{\mu\nu}-\frac{q^\mu q^\nu}{q^\alpha q_\alpha}\biggr)\,.
\ee
As a result, in the vacuum limit where $\Pi_{\mu\nu} = \bigl(g_{\mu\nu} - q_\mu q_\nu/(q^\alpha q_\alpha)\bigr)\Pi$ and the photon propagator is proportional to $\frac{1}{q^\alpha q_\alpha-\Pi}$, we have ${\cal K}_{\lambda\lambda'}=\Pi\,\delta_{\lambda\lambda'}$. In an isotropic medium,
\be
\Pi^{\mu\nu} = -\Pi_T \sum_{\lambda=\pm} \epsilon_\lambda^\mu \epsilon_\lambda^{\nu*} -\Pi_L \epsilon_L^\mu \epsilon_L^{\nu*}\,,\qquad
{\cal K} = \text{diag}(\Pi_T,\, \Pi_T,\, \Pi_L)\,,
\ee
and the photon propagators are proportional to $\frac{1}{q^\alpha q_\alpha-\Pi_{T,L}}$. Generically, for an anisotropic medium, we need to simultaneously rotate $A$ and $A'$ into a polarization basis where ${\cal K}$ is diagonal. In this basis, the quadratic part of the effective action can be diagonalized for each polarization by
\be
A_\mu = \widetilde A_\mu +\kappa\frac{\Pi}{m_{A'}^2-\Pi} \widetilde A'_\mu\,,\qquad
A'_\mu = \widetilde A'_\mu -\kappa\frac{\Pi}{m_{A'}^2-\Pi} \widetilde A_\mu\,,
\ee
where $\Pi$ is an eigenvalue of ${\cal K}$. In the $\widetilde A, \widetilde A'$ basis, the propagators are proportional to $\frac{1}{q^\alpha q_\alpha-\Pi}$ and $\frac{1}{q^\alpha q_\alpha-m_{A'}^2}$, respectively, and the interactions in Eq.~\eqref{eq:Lvac} read 
\beq
&& \biggl[ e (J_p^\mu -J_e^\mu) -\frac{\Pi}{m_{A'}^2-\Pi}\kappa g_\chi J_\chi^\mu \biggr] \widetilde A_\mu \nonumber\\
&& + \biggl[ g_\chi J_\chi^\mu + \biggl(f_p^0-\frac{\Pi}{m_{A'}^2-\Pi}f_e^0\biggr) J_p^\mu +f_n^0 J_n^\mu +\frac{m_{A'}^2}{m_{A'}^2-\Pi}f_e^0 J_e^\mu\biggr] \widetilde A'_\mu\,.
\eeq
Dark matter scattering is mediated by both $\widetilde A$ and $\widetilde A'$. Taking both into account, we obtain the following effective interaction:
\beq
&& g_\chi J_{\chi\mu} \biggl\{ -\frac{1}{q^\alpha q_\alpha-\Pi}\frac{\Pi}{m_{A'}^2-\Pi} \kappa e(J_p^\mu - J_e^\mu) \nonumber\\
&&\qquad\quad +\frac{1}{q^\alpha q_\alpha-m_{A'}^2} \biggl[ \Bigl(f_p^0-\frac{\Pi}{m_{A'}^2-\Pi}f_e^0\Bigr) J_p^\mu +f_n^0 J_n^\mu +\frac{m_{A'}^2}{m_{A'}^2-\Pi}f_e^0 J_e^\mu\biggr]\biggr\} \nonumber\\[4pt]
&=& \frac{1}{q^\alpha q_\alpha-m_{A'}^2} g_\chi J_{\chi\mu} \biggl\{ \biggl[ f_p^0 + \Bigl(1-\frac{q^\alpha q_\alpha}{q^\alpha q_\alpha-\Pi}\Bigr)f_e^0\biggr] J_p^\mu + f_n^0 J_n^\mu + \frac{q^\alpha q_\alpha}{q^\alpha q_\alpha-\Pi} f_e^0 J_e^\mu\biggr\} \label{eq:imeffint0}\\
&=& \frac{1}{q^\alpha q_\alpha-m_{A'}^2} g_\chi J_{\chi\mu} \biggl[ \frac{q^\alpha q_\alpha}{q^\alpha q_\alpha-\Pi} f_e^0 (J_e^\mu - J_p^\mu) +(f_p^0+f_e^0) J_p^\mu +f_n^0 J_n^\mu \biggr]
\label{eq:imeffint}
\eeq
From the last equation, it is clear that the current $A'$ couples to contains a screened component and an unscreened component: $f_p^0 J_p^\mu + f_n^0 J_n^\mu + f_e^0 J^\mu_e = f_e^0(J_e^\mu-J_p^\mu) + \bigl[(f_p^0+f_e^0) J_p^\mu +f_n^0 J_n^\mu\bigr]$. The first term, which is proportional to the electromagnetic current, gets screened by a factor of $\frac{q^\alpha q_\alpha}{q^\alpha q_\alpha-\Pi}$, whereas the second term is unaffected.

In the special case of a dark photon that kinetically mixes with the SM photon, Eq.~\eqref{eq:Lvac} follows from diagonalizing the kinetic terms, and $\kappa$ is equal to the kinetic mixing parameter. In this case, $f_p^0 = -f_e^0 = \kappa e, f_n^0 = 0$, and the DM interaction is maximally screened. In contrast, a $U(1)_{B-L}$ gauge boson has $f_p^0 = f_n^0 = -f_e^0$, and the coupling to neutrons is not screened. As a final example, a hadrophobic $A'$ has $f_p^0 = f_n^0 = 0$, resulting in an unscreened DM coupling to protons (which originates from the $A$-$A'$ mixing).

The screening factor $\frac{q^\alpha q_\alpha}{q^\alpha q_\alpha-\Pi}$ can be expressed in terms of the dielectric matrix $\vect{\varepsilon}(\vect{q},\omega)$ by solving the following set of equations for $\Pi^{\mu\nu}$~\cite{Hochberg:2015fth,Coskuner:2019odd}:
\beq
J^\mu &=& -\Pi^{\mu\nu} A_\nu\,,\\
J^i &=& {\sigma^i}_j E^j = {\sigma^i}_j (i\omega A^j -i q^j A^0) \,,\\
\vect{\sigma} &=& \vect{\sigma}^T = i\omega(\mathbbm{1}-\vect{\varepsilon})\,.
\eeq
Note that the three-dimensional quantities are defined by $\vect{\sigma} = {\sigma^i}_j$, $\mathbbm{1} = {\delta^i}_j$, $\vect{\varepsilon} = {\varepsilon^i}_j$. We obtain the following solution:
\bea
&& \Pi^{\mu\nu} = \left(
\begin{matrix}
\Pi_{00} & \vect{\Pi}_0 \\
\vect{\Pi}_0 & -\vect{\Pi}
\end{matrix}
\right)\,, \\
\Pi_{00} = \frac{i}{\omega} \vect{q}\cdot\vect{\sigma}\cdot\vect{q}\,,\qquad
&& \vect{\Pi}_0 \equiv {\Pi_0}^i = i\vect{\sigma}\cdot\vect{q}\,,\qquad
\vect{\Pi} \equiv {\Pi^i}_j = -i\omega\vect{\sigma}\,.
\eeq
Projecting $\Pi^{\mu\nu}$ onto polarization components, we obtain:
\begin{align}
{\cal K}_{LL} &= q^\alpha q_\alpha (1-\vect{\hat q}\cdot\vect{\varepsilon}\cdot\vect{\hat q})\,,&
{\cal K}_{L\pm} &= {\cal K}_{L\mp} = -\omega\sqrt{q^\alpha q_\alpha} \,\vect{\hat q}\cdot\vect{\varepsilon}\cdot\vect{\epsilon}_\pm \,,\\
{\cal K}_{\pm\pm} &= \omega^2 \bigl(1-\vect{\epsilon}_\mp\cdot\vect{\varepsilon}\cdot\vect{\epsilon}_\pm\bigr) \,,&
{\cal K}_{\mp\pm} &= -\omega^2 \,\vect{\epsilon}_\pm\cdot\vect{\varepsilon}\cdot\vect{\epsilon}_\pm \,.
\end{align}
We can see explicitly that in the isotropic limit, $\vect{\varepsilon}\propto\mathbbm{1}$, so ${\cal K}_{L\pm} = {\cal K}_{\mp\pm} = 0$, and ${\cal K}_{LL}$, ${\cal K}_{\pm\pm}$ are identified as $\Pi_L$, $\Pi_T$, respectively. In this case, ${\cal K} = \text{diag}(\Pi_L,\,\Pi_T,\,\Pi_T)$, and the familiar relations
\be
\Pi_L = q^\alpha q_\alpha(1-\varepsilon),\,\quad \Pi_T=\omega^2(1-\varepsilon) 
\ee
are reproduced. Beyond the isotropic limit, in general one has to diagonalize the ${\cal K}$ matrix as discussed above. However, assuming anisotropies are not large, the calculation is simplified in the case of nonrelativistic scattering. Here, the currents involved ($J_\chi^\mu$, $J_e^\mu$, etc.)\ have velocity suppressed spatial components, so the dominant contribution comes from the polarization that is almost longitudinal, for which $\Pi\simeq {\cal K}_{LL}$ up to small corrections. As a result, the screening factor in Eq.~\eqref{eq:imeffint} becomes
\be
\frac{q^\alpha q_\alpha}{q^\alpha q_\alpha-\Pi} \simeq \frac{q^2}{\vect{q}\cdot\vect{\varepsilon}\cdot\vect{q}}\,.
\ee
Now it is straightforward to read off the screening of DM couplings from Eq.~\eqref{eq:imeffint0}:
\be
f_p(\vect{q}) = f_p^0+\biggl(1-\frac{q^2}{\vect{q}\cdot\vect{\varepsilon}\cdot\vect{q}}\biggr)\,f_e^0\,,\qquad
f_n(\vect{q}) = f_n^0 \,,\qquad
f_e(\vect{q}) = \frac{q^2}{\vect{q}\cdot\vect{\varepsilon}\cdot\vect{q}}\,f_e^0\,.
\label{eq:screeningfactors}
\ee
In what follows, we will often drop the argument $\vect{q}$ and just write $f_p, f_n, f_e$ for simplicity. Finally, we note that in the scattering limit, $q\gg\omega$, a scalar mediator has the same coupling as the longitudinal component of a vector mediator, so the same screening factors in Eq.~\eqref{eq:screeningfactors} apply.

To close this subsection, we comment that in-medium screening affects different channels differently. Nuclear recoils happen at high enough momentum transfer where $\vect{\varepsilon}$ can be approximated as unity, so $f_\psi\simeq f_\psi^0$. For electron transitions, the situation depends on the band gap. For atoms, insulators and semiconductors with $\OO(\text{eV})$ or larger band gaps, $\vect{\varepsilon}$ approaches unity when $q\gtrsim 2 \pi/a\sim \mathcal{O}(\text{keV})$~\cite{Cappellini_et_al:1993}; for smaller $q$, the full $\vect{\varepsilon} (q)$ can be fitted to experimental measurements or calculated using advanced electronic structure techniques. For small-gap systems such as superconductors and Dirac semi-metals, it is important to keep the full energy-momentum dependence in $\vect{\varepsilon}(\vect{q},\omega)$. For example, in a (super)conductor, $\vect{\varepsilon}\sim \lambda_\text{TF}^2/q^2$ at low $q$, where $\lambda_\text{TF}\sim\OO(\text{keV})$ is the Thomas-Fermi screening parameter, resulting in significant screening~\cite{Hochberg:2015fth}. In contrast, in a Dirac semi-metal, $\vect{\varepsilon}$ approaches a constant at low $q$, so sensitivity to dark photon mediated scattering (and also dark photon absorption) is much stronger~\cite{Hochberg:2017wce,Coskuner:2019odd}. For phonon excitations, screening from electrons should also be accounted for, as we discuss in Sec.~\ref{sec:phonon-e}. 

\section{Nuclear Recoils}
\label{sec:nuclear}

We now apply the general framework of the previous section to the case of nuclear recoils and reproduce familiar results. For simplicity we shall first assume only one type of nucleus is present, with proton number $Z$ and atomic mass number $A$, and later generalize to the case of multiple nucleus types with non-degenerate $\{Z_N\}, \{A_N\}$.

To begin, we assume the nuclei do not interact with each other, so the Hilbert space of the target system, which contains $\rho_T V/m_N$ nuclei, is a direct product of $\rho_T V/m_N$ single nucleus Hilbert spaces. We will discuss the validity of this standard assumption in Sec.~\ref{sec:nuclear-validity}. The target system is prepared in the initial state
\be
|i\rangle= \prod_{J=1}^{\rho_T V/m_N} |\vect{k}_i\rangle_J = |\vect{k}_i\rangle_1\otimes|\vect{k}_i\rangle_2\otimes\dots
\ee
with $\vect{k}_i = \vect{0}$. In the final state $|f\rangle$, one of the $|\vect{k}_i\rangle_J$'s is replaced by $|\vect{k}_f\rangle_J$ with $\vect{k}_f \neq \vect{0}$. We can write these states in terms of nucleus creation operators:
\be
|\vect{k}_i\rangle_J=V^{-1/2}\, \hat b_{\vect{k}_i}^\dagger|0\rangle_J\,,\qquad
|\vect{k}_f\rangle_J=V^{-1/2}\, \hat b_{\vect{k}_f}^\dagger|0\rangle_J\,.
\ee
As usual, we have the canonical commutation relations $[\hat b_{\vect{k}}, \hat b_{\vect{k}'}^\dagger] = (2\pi)^3\delta^3(\vect{k}-\vect{k}')$ or $\{\hat b_{\vect{k}}, \hat b_{\vect{k}'}^\dagger\} = (2\pi)^3\delta^3(\vect{k}-\vect{k}')$, etc.

Now we need to quantize
\be
{\cal F}_T (\vect{q}) = \frac{1}{f_n}\bigl[ f_p \widetilde n_p(-\vect{q}) +f_n \widetilde n_n(-\vect{q}) +f_e \widetilde n_e(-\vect{q}) \bigr]
\ee
in terms of nucleus creation and annihilation operators $\hat b^\dagger, \hat b$. Obviously, the electron coupling does not contribute, so we drop the last term. The proton and neutron number densities, on the other hand, can be related to the nucleus number density $n_N$, if we assume elastic scattering (no transition between nuclear states):
\be
n_{p,n}(\vect{x}') = \int d^3 x''\, n_N(\vect{x}'')\, n_{p,n}^0(\vect{x}'-\vect{x}'')\,,
\ee
where $n_{p,n}^0$ are the proton and neutron number densities around a single nucleus at the origin. Therefore,
\be
{\cal F}_T(\vect{q}) = \frac{f_p \widetilde n_p^0(-\vect{q}) +f_n \widetilde n_n^0(-\vect{q})}{f_n} \,\widetilde n_N(-\vect{q}) \equiv \frac{f_N}{f_n} F_N(\vect{q})\,\widetilde n_N(-\vect{q})\,,
\ee
where $f_N\equiv f_p Z+f_n(A-Z)$, the DM-nucleus coupling in the $q\to0$ limit (where DM interacts with all nucleons coherently). $F_N(\vect{q})$ is a nuclear form factor that deviates from unity only for $q$ above the inverse nucleus radius. A commonly used form factor is the Helm form factor~\cite{Helm:1956zz},
\be
F_N(\vect{q}) = \frac{3\,j_1(qr_n)}{qr_n}\,e^{-(qs)^2/2} = 1-\frac{(qr_n)^2}{10}-\frac{(qs)^2}{2} +\OO(q^4)\,, \label{eq:helm}
\ee
where $r_n \simeq 1.14\, A^{1/3} \,\text{fm}$, $s\simeq 0.9 \,\text{fm}$. We can thus write ${\cal F}_T(\vect{q})$ in terms of $\hat b^\dagger, \hat b$ via
\be
\widetilde n_N(-\vect{q}) = \int d^3x \,e^{i\vect{q}\cdot\vect{x}}\,\hat b_{\vect{x}}^\dagger \hat b_{\vect{x}}
= \int \frac{d^3k'}{(2\pi)^3} \frac{d^3k}{(2\pi)^3} \,(2\pi)^3\delta^3(\vect{k}'-\vect{k}-\vect{q}) \,\hat b_{\vect{k}'}^\dagger \hat b_{\vect{k}}\,.
\ee

To obtain the dynamic structure factor, we evaluate the matrix element,
\beq
_J\langle \vect{k}_f| \widetilde n_N(-\vect{q}) |\vect{k}_i\rangle_J &=& \frac{1}{V} \int \frac{d^3k'}{(2\pi)^3} \frac{d^3k}{(2\pi)^3} \,(2\pi)^3\delta^3(\vect{k}'-\vect{k}-\vect{q}) \,\langle 0|\, \hat b_{\vect{k}_f} \hat b_{\vect{k}'}^\dagger \hat b_{\vect{k}} \hat b_{\vect{k}_i}^\dagger |0\rangle \nonumber\\
&=& \frac{(2\pi)^3}{V} \delta^3(\vect{k}_f-\vect{k}_i-\vect{q})\,,
\eeq
and sum over final states, which amounts to summing over the scattered nucleus $J$ (simply multiplying by $\rho_T V/m_N$) and integrating over the final momentum $V\int d^3k_f/(2\pi)^3$. Therefore,
\beq
S(\vect{q},\omega) 
&=& 2 \pi \frac{\rho_T}{m_N} \frac{f_N^2}{f_n^2} \,F_N^2(\vect{q})\cdot V\int \frac{d^3k_f}{(2\pi)^3} \biggl[\frac{(2\pi)^3}{V} \delta^3(\vect{k}_f-\vect{k}_i-\vect{q})\biggr]^2 \,\delta \biggl( \omega - \frac{q^2}{2m_N} \biggr)
\nonumber\\
&=& 2 \pi \frac{\rho_T}{m_N} \frac{f_N^2}{f_n^2} \,F_N^2(\vect{q})\,\delta \biggl( \omega - \frac{q^2}{2m_N} \biggr),
\eeq
where we have regulated the delta function by $\frac{(2\pi)^3}{V}\delta^3(\vect{0})=\frac{1}{V}\int d^3x \,e^{i\vect{0}\cdot\vect{x}}=1$.

We can now reproduce the familiar results for the differential rate. Assuming the nuclear form factor is isotropic, $F_N(\vect{q})=F_N(q)$, as is the case for the Helm form factor in Eq.~\eqref{eq:helm}, we can apply Eq.~\eqref{eq:R-iso} and obtain
\beq
\frac{dR}{d\omega} &=& \frac{\rho_\chi}{m_\chi}\frac{\overline\sigma_n}{2\mu_{\chi n}^2} \frac{f_N^2}{f_n^2} \int dq\, F_N^2\, {\cal F}_\text{med}^2\, \eta(v_\text{min})\, \frac{q}{m_N}\,\delta \biggl( \omega - \frac{q^2}{2m_N} \biggr) \label{eq:dRdED-nr-1}\\
&=& \frac{\rho_\chi}{m_\chi}\frac{\overline\sigma_n}{2\mu_{\chi n}^2}\frac{f_N^2}{f_n^2}\,F_N^2\, {\cal F}_\text{med}^2\,\eta( v_\text{min})\Bigr|_{q^2=2m_N\omega}\,,
\eeq
where $\eta(v_\text{min})$ is given by Eq.~\eqref{eq:etafun} and $v_\text{min}=\frac{q}{2\mu_{\chi N}}$ in the present case. It is now easy to generalize these results to the case of more than one nucleus type:
\be
\frac{dR}{d\omega} = \frac{\rho_\chi}{m_\chi}\frac{\overline\sigma_n}{2\mu_{\chi n}^2} \frac{1}{\sum_N A_N} \biggl[\sum_N A_N\, \frac{f_N^2}{f_n^2}\,F_N^2\,{\cal F}_\text{med}^2 \,\eta (v_{\text{min}})\biggr]_{q^2=2m_N \omega}\,,\label{eq:dRdED-nr-2}
\ee
where $N$ runs over the inequivalent nuclei in the target (e.g.\ $N=$ Ga,\,As for GaAs).

\subsection{Validity of the Nuclear Recoil Calculation in Crystal Targets}
\label{sec:nuclear-validity}

A key assumption we have made in the derivation above is that the nuclei in the target do not interact with each other (hence the factorization of the Hilbert space). In a crystal target, however, the nuclei are not free, but interact with the neighboring nuclei in the crystal structure. The justification of treating the nuclei as free particles initially at rest lies in the fact that in the {\it classical}  limit, the hard scattering process is instantaneous and local. In this case, the nuclei interactions affect only the subsequent secondary processes. For example, secondary phonons can be produced, which allows the energy deposition to be shared by many nuclei. 

On the other hand, as detector thresholds are pushed to lower energies, at some point we would get into the {\it quantum} regime, where the finite duration and spatial extent of the scattering invalidate the free nuclei assumption. We can make a quick estimate on when this happens from the uncertainty principle. 
The time scale for the hard scattering to happen is $\sim 1/\omega$. This should be compared to the intrinsic time scale for atomic vibrations in a crystal, $1/\omega_\text{ph}$, with $\omega_\text{ph}$ the phonon energy. The instantaneous interaction approximation in the standard nuclear recoil calculation is valid when the energy deposition is much higher than the energies of all phonon modes, i.e.
\be
\omega\gg\omega_\text{ph}^\text{max}  \qquad\quad\text{(validity condition for nuclear recoils in crystals)}\,.
\label{eq:nrvalidity}
\ee

An alternative way to reach the same conclusion is the following. Within the length scale $1/q$, the DM should see the nucleus as a plane wave for the nuclear recoil calculation to hold. Since the spatial extent of the nucleus wavefunction in a harmonic potential is $\sim (m_N\omega_\text{ph})^{-1/2}$, we need $q\gg (m_N\omega_\text{ph}^\text{max})^{1/2}$. Using the kinematic relation $\omega=\frac{q^2}{2m_N}$, we arrive at the same condition as Eq.~\eqref{eq:nrvalidity}.

To summarize, in crystal targets, the nuclear recoil calculation is valid for energy depositions much higher than the phonon energies, which are typically $\OO(10 - 100)$ meV. This explains the truncation of the C, Si, Ge, Cs nuclear recoil lines at low $\omega$ in Fig.~\ref{fig:kin}. At lower energy depositions, the target Hilbert space does not factorize into individual nuclei, but instead contains single phonon and multi-phonon states as energy eigenstates, and the direct detection rate calculation proceeds differently. We discuss single phonon excitations in Sec.~\ref{sec:phonon}, which will be the relevant processes when detector thresholds reach the 10-100 meV regime in the future. In the intermediate energy regime -- above the single phonon energies yet below the validity range of nuclear recoils -- direct multi-phonon production should be considered, which we plan to investigate in future work.

\section{Electron Transitions}
\label{sec:electron}

We next consider electron transitions. The initial state can be written as
\be
|i\rangle = \prod_{I\in\,\text{occupied}} \hat c_I^\dagger\, |0\rangle\,,
\ee
where $\hat c_I^\dagger$ are electron creation operators, with $I$ running over all occupied electron states (energy eigenstates). Our normalization convention is such that $\{\hat c_{I},\, \hat c_{I'}^\dagger\} = \delta_{II'}$, so the electron states are unit-normalized. The final states are labeled by $I_1, I_2$, where one of the electrons has transitioned from $I_1$ to an unoccupied state $I_2$:
\be
|f\rangle = \hat c_{I_2}^\dagger \hat c_{I_1} |i\rangle\,.
\ee

The relevant piece in ${\cal F}_T(\vect{q})$ is simply
\be
{\cal F}_T(\vect{q}) = \frac{f_e}{f_e^0}\, \widetilde n_e(-\vect{q}) = \frac{f_e}{f_e^0} \int \frac{d^3k'}{(2\pi)^3} \frac{d^3k}{(2\pi)^3} \,(2\pi)^3\delta^3(\vect{k}'-\vect{k}-\vect{q}) \,\hat c_{\vect{k}'}^\dagger \hat c_{\vect{k}}\,,
\ee
where the creation and annihilation operators are for momentum eigenstates, and satisfy $\{\hat c_{\vect{k}}, \hat c_{\vect{k}'}^\dagger\} = (2\pi)^3\delta^3(\vect{k}-\vect{k}')$, etc. As discussed in Sec.~\ref{sec:general-medium}, the screening factor is
\be
\frac{f_e}{f_e^0} = 
\frac{q^2}{\vect{q}\cdot\vect{\varepsilon}\cdot\vect{q}}
\label{eq:feonfe0}
\ee
for a vector or scalar mediator.

The dynamic structure factor is therefore
\beq
S\bigl(\vect{q},\omega\bigr) 
&=& \frac{2 \pi }{V} \biggl(\frac{f_e}{f_e^0}\biggr)^2 \sum_{I_1,I_2} \,\delta\bigl(E_{I_2}-E_{I_1}-\omega\bigr) \times\nonumber\\
&& \qquad\qquad\qquad  \biggl|\int \frac{d^3k'}{(2\pi)^3} \frac{d^3k}{(2\pi)^3} \,(2\pi)^3\delta^3(\vect{k}'-\vect{k}-\vect{q}) \langle i| \hat c_{I_1}^\dagger \hat c_{I_2} \hat c_{\vect{k}'}^\dagger \hat c_{\vect{k}}  |i\rangle \biggr|^2 \nonumber\\
&=& \frac{2 \pi}{V} \biggl(\frac{f_e}{f_e^0}\biggr)^2 \sum_{I_1,I_2} \, \delta\bigl(E_{I_2}-E_{I_1}-\omega\bigr) \times\nonumber\\
&&\qquad\qquad\qquad \biggl| \int \frac{d^3k'}{(2\pi)^3} \frac{d^3k}{(2\pi)^3} \,(2\pi)^3\delta^3(\vect{k}'-\vect{k}-\vect{q}) \,\{ \hat c_{\vect{k}},\, \hat c_{I_1}^\dagger\} \{\hat c_{I_2},\, \hat c_{\vect{k}'}^\dagger \} \biggr|^2,\quad
\eeq
where we have used $\hat c_{I_1}^\dagger|i\rangle = \hat c_{I_2} |i\rangle = 0$, and that the anticommutators are just numbers. To evaluate the anticommutators, we expand the energy eigenstates in terms of momentum eigenstates:
\be
\hat c_I^\dagger |0\rangle = \int \frac{d^3k}{(2\pi)^3} \,\widetilde\psi_I(\vect{k})\, \hat c_{\vect{k}}^\dagger|0\rangle\,,
\ee
where $\widetilde\psi_I(\vect{k})$ is the momentum space wavefunction, which satisfies the orthonormality condition $\int \frac{d^3k}{(2\pi)^3} \,\widetilde\psi_{I'}^*(\vect{k}) \widetilde\psi_I(\vect{k}) =\delta_{II'}$. We then obtain
\beq
S\bigl(\vect{q},\omega\bigr) 
&=& \frac{2 \pi}{V} \biggl(\frac{f_e}{f_e^0}\biggr)^2 \sum_{I_1,I_2}\,\delta\bigl(E_{I_2}-E_{I_1}-\omega\bigr) \cdot \nonumber\\
&&\qquad\qquad\qquad \biggl| \int \frac{d^3k'}{(2\pi)^3} \frac{d^3k}{(2\pi)^3} \,(2\pi)^3\delta^3(\vect{k}'-\vect{k}-\vect{q}) \,\widetilde\psi_{I_2}^* (\vect{k}')\widetilde\psi_{I_1} (\vect{k}) \biggr|^2 .
\label{eq:Se}
\eeq

The dynamic structure factor in Eq.~\eqref{eq:Se} applies for any target system where DM scattering can trigger electron transitions -- atoms, crystals, superconductors, Dirac materials, etc.\ -- once the energy levels and wavefunctions are known. In what follows, we examine the case of periodic crystals in more detail. Here, the energy eigenstates of an electron are Bloch waves labeled by a band index and a wavevector within the first Brillouin zone (1BZ), e.g.
\beq
\psi_{I_1}(\vect{x}) &=& \psi_{i_1\vect{k}_1} (\vect{x}) = \frac{1}{\sqrt{V}} \sum_{\vect{G}_1} u_{i_1}(\vect{k}_1+\vect{G}_1)\,e^{i(\vect{k}_1+\vect{G}_1)\cdot\vect{x}}\,,\\
\widetilde\psi_{i_1\vect{k}_1}(\vect{k}) &=& \int d^3x \,\psi_{i_1\vect{k}_1} (\vect{x})\,e^{-i\vect{k}\cdot\vect{x}} = \frac{1}{\sqrt{V}} \sum_{\vect{G}_1} u_{i_1}(\vect{k}_1+\vect{G}_1)\,(2\pi)^3\delta^3(\vect{k}_1+\vect{G}_1-\vect{k})\,,\quad
\eeq
where $\vect{G}_1$ runs over all reciprocal lattice vectors. Note that the state labeled by $i_1,\vect{k}_1$ has Fourier components of $\vect{k}_1$ plus any reciprocal lattice vector. The coefficients $u_{i_1}(\vect{k}_1+\vect{G}_1)$ are normalized as $\sum_{\vect{G}_1} |u_{i_1}(\vect{k}_1+\vect{G}_1)|^2=1$. The dynamic structure factor now becomes
\beq
S\bigl(\vect{q},\omega\bigr) 
&=& \frac{2}{V} \biggl(\frac{f_e}{f_e^0}\biggr)^2 \sum_{i_1,i_2} \int_\text{1BZ} \frac{d^3k_1}{(2\pi)^3} \frac{d^3k_2}{(2\pi)^3}\, 2\pi\,\delta\bigl(E_{i_2,\vect{k}_2}-E_{i_1,\vect{k}_1}-\omega\bigr) \times\nonumber\\
&&\qquad \biggl| \sum_{\vect{G}_1,\vect{G}_2} \,(2\pi)^3\delta^3(\vect{k}_2+\vect{G}_2-\vect{k}_1-\vect{G}_1-\vect{q}) \,u_{i_2}^*(\vect{k}_2+\vect{G}_2)\, u_{i_1}(\vect{k}_1+\vect{G}_1) \biggr|^2 ,\qquad
\eeq
where the prefactor 2 comes from summing over contributions from degenerate spin states, and the sums over the final state quantum numbers $\vect{k}_{1,2}$ have been replaced by integrals in the continuum limit. As in Ref.~\cite{Essig:2015cda}, we define a crystal form factor
\be
f_{[i_1\vect{k}_1,i_2\vect{k}_2,\vect{G}]} \equiv \sum_{\vect{G}_1,\vect{G}_2} u_{i_2}^* \bigl(\vect{k}_2+\vect{G}_2 \bigr) u_{i_1} \bigl(\vect{k}_1+\vect{G}_1 \bigr) \,\delta_{\vect{G}_2-\vect{G}_1,\vect{G}}
\label{eq:fcrystal}
\ee
for the transition $i_1\vect{k}_1\to i_2\vect{k}_2$ with an Umklapp $\vect{G}$. This simply encodes the wavefunction overlap, summed over all Fourier components consistent with momentum conservation. The dynamic structure factor can now be written more concisely as
\beq
S\bigl(\vect{q},\omega\bigr) 
&=& 2\biggl(\frac{f_e}{f_e^0}\biggr)^2  \sum_{i_1,i_2} \int_\text{1BZ} \frac{d^3k_1}{(2\pi)^3} \frac{d^3k_2}{(2\pi)^3} \,2\pi\,\delta\bigl(E_{i_2,\vect{k}_2}-E_{i_1,\vect{k}_1}-\omega\bigr)\times \nonumber\\
&&\qquad\qquad\quad \sum_{\vect{G}}\, (2\pi)^3\delta^3(\vect{k}_2-\vect{k}_1+\vect{G}-\vect{q}) \,\bigl|f_{[i_1\vect{k}_1,i_2\vect{k}_2,\vect{G}]}\bigr|^2 .\quad
\eeq
Note that we have again used the identity $(2\pi)^3\delta^3(\vect{0})=\int d^3x \,e^{i\vect{0}\cdot\vect{x}}=V$. The material-specific quantities appearing in $S(\vect{q},\omega)$ are the electron band structures (energy eigenvalues $E_{i,\vect{k}}$) and Bloch wavefunction coefficients $u_{i}(\vect{k}+\vect{G})$. They can be computed by density functional theory (DFT) methods which we discuss more in our companion paper~\cite{Griffin:2019mvc}.

Finally, performing the phase space integration, we obtain the total rate per target mass:
\be
R = \frac{2}{\rho_T} \frac{\rho_\chi}{m_\chi} \frac{\pi\overline\sigma_e}{\mu_{\chi e}^2} 
\sum_{i_1,i_2} \int_\text{1BZ} \frac{d^3k_1}{(2\pi)^3} \frac{d^3k_2}{(2\pi)^3} \sum_{\vect{G}} g(\vect{q},\omega)\, {\cal F}_\text{med}^2(q)\biggl(\frac{f_e}{f_e^0}\biggr)^2\bigl|f_{[i_1\vect{k}_1,i_2\vect{k}_2,\vect{G}]}\bigr|^2 \,,
\label{eq:rate-e}
\ee
where
\be
\vect{q} = \vect{k}_2-\vect{k}_1 +\vect{G}\,,\qquad
\omega = E_{i_2,\vect{k}_2} -E_{i_1,\vect{k}_1}\,.
\ee
The $g(\vect{q}, \omega)$ function, the mediator form factor ${\cal F}_\text{med}$, the screening factor $f_e/f_e^0$ and the crystal form factor $f_{[i_1\vect{k}_1,i_2\vect{k}_2,\vect{G}]}$ are given by Eqs.~\eqref{eq:gfun}, \eqref{eq:Fmed}, \eqref{eq:feonfe0} and \eqref{eq:fcrystal}, respectively. This generalizes the formula derived in Ref.~\cite{Essig:2015cda} to account for possible anisotropies in the target response.

\subsection{Target Anisotropies and Daily Modulation}
\label{sec:electron-modulation}

The simplest crystal targets that have been considered for direct detection via electron transitions, like silicon and germanium, are quite isotropic. As a result, the rate is essentially independent of the direction of the incoming DM's velocity. However, this is not the case for materials with large anisotropies in the electron band structures or wavefunctions. For terrestrial experiments, as the target rotates with the Earth, the DM wind comes in from different directions at different times of the day, resulting in a {\it daily} modulation of the rate. This is on top of the {\it annual} modulation signal expected due to the variation of the average DM velocity as the Earth orbits around the Sun~\cite{Lee:2015qva,Essig:2015cda}. If observed, it would be a smoking-gun signature of DM that is distinct from possible backgrounds. Our rate formula Eq.~\eqref{eq:rate-e} incorporates directional information, and is well-suited for calculating the daily modulation signal.

\begin{figure}[t]
\includegraphics[width=\linewidth]{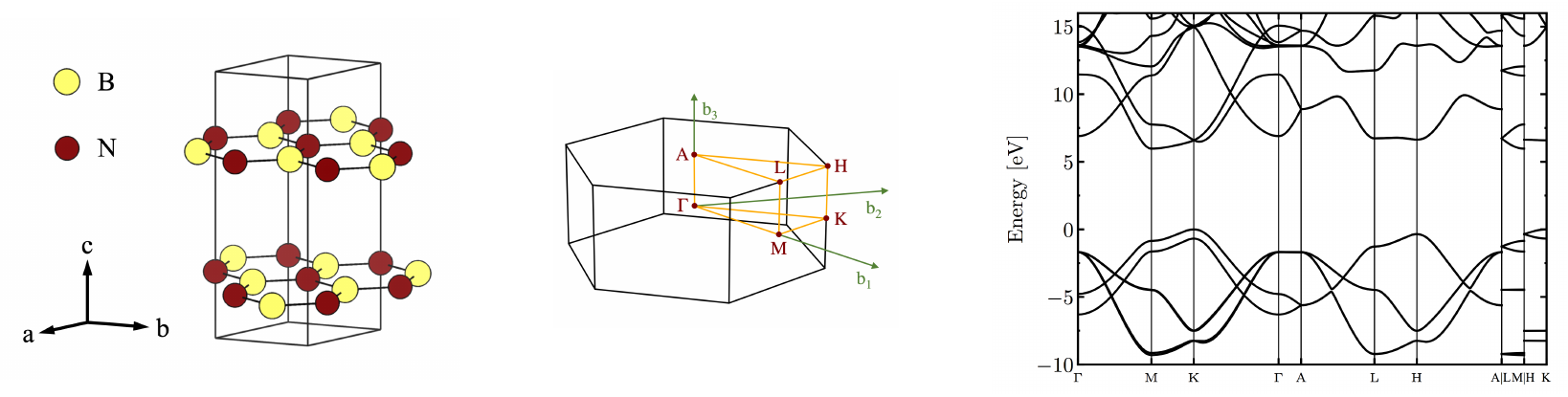}
\caption{\label{fig:bn}
Crystal structure of hexagonal boron nitride (left), its corresponding first Brillouin zone (middle) and DFT-calculated electronic band structure (right) with the Fermi level set to zero. The letters shown in the Brillouin zone plot mark several of the high-symmetry points, and the orange lines mark Brillouin zone paths along which electronic band structure is plotted. The discontinuities in the band structure occur from taking a discontinuous path through the Brillouin zone, indicated by ``\,$|$\,'' on the horizontal axis.
}
\end{figure}

As an example target, we consider hexagonal boron nitride (BN), shown in Fig.~\ref{fig:bn}. The numerical calculation of electron band structures and wavefunction coefficients, as well as direct detection rates, proceeds in the same way as in our companion paper~\cite{Griffin:2019mvc}. We include the calculation details specific for BN in Appendix~\ref{app:dft}. As a result of the layered crystal structure, the rate is strongly dependent on the angle between the DM wind and the layers. We note, however, that BN has a three-dimensional crystal structure with the layers of BN repeating in the out-of-plane direction, in contrast to single-layer graphene previously considered in Ref.~\cite{Hochberg:2016ntt}.

\begin{figure}[t]
\includegraphics[width=\linewidth]{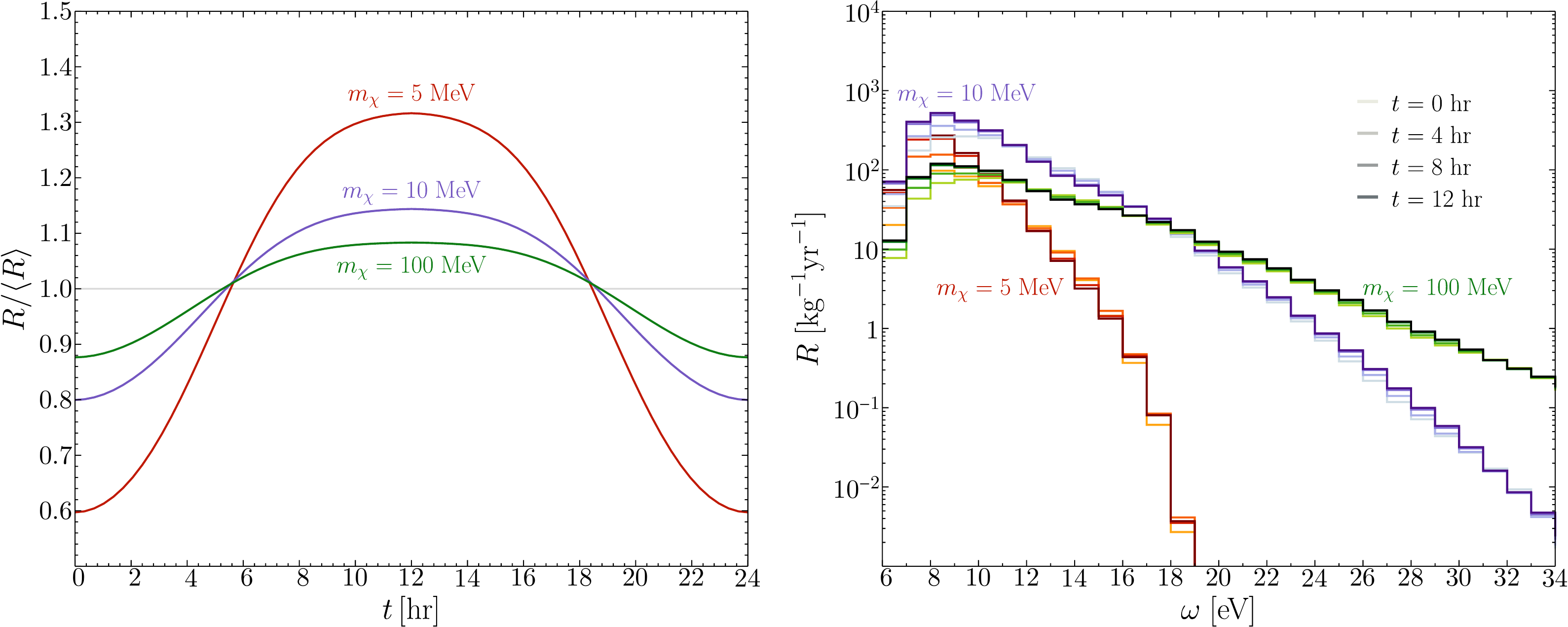}
\caption{\label{fig:mod}
Total rate of electron transitions $R$ in hexagonal BN, normalized to its daily average $\langle R\rangle$ as a function of time (left), and differential rates at several times of the day assuming $\overline{\sigma}_e=10^{-37}\,\text{cm}^2$ (right), for a $5, 10, 100\,$MeV DM scattering via a light mediator, neglecting screening.
}
\end{figure}

To show this directional dependence, we consider the same experimental setup as in Refs.~\cite{Griffin:2018bjn,Coskuner:2019odd}, where the crystal $c$-axis is aligned with the Earth's velocity $\vect{v}_e$ at time $t=0$. With this choice, daily modulation signal is independent of the location of the laboratory. In Fig.~\ref{fig:mod}, we pick three DM masses $m_\chi = 5, 10, 100\,$MeV to show how the expected detection rates -- both total (left panel) and differential (right panel) -- change during a sidereal day, assuming a light mediator and negligible in-medium effects. For all three masses, we see that the rate is maximized at $t=12\,$hours when the DM wind is roughly aligned with the crystal $a$-$b$ plane, and minimized at $t=0$ when the DM wind is aligned with the crystal $c$-axis. This can be understood from the fact that electron wavefunctions are more localized in the $c$ direction and thus have smaller low-momentum components, whereas the DM scattering matrix element peaks at low $q$ for a light mediator. We also observe that modulation is stronger for lighter DM. Generically, with a smaller energy deposition, the rate is more strongly affected by band structure anisotropies near the band gap; far from the band gap, the electron band structures and wavefunctions approach those for individual, isotropic ions. For DM heavier than 100\,MeV, we find roughly the same amount of daily modulation as the $m_\chi=100\,$MeV case. This is again because the momentum integral is dominated by small $q$, which corresponds to the same kinematic region $\omega_{\vect{q}} \simeq \vect{q}\cdot\vect{v}$ in the large $m_\chi$ limit. On the other hand, once we go below $m_\chi=5\,$MeV, the total rate quickly approaches zero, as the DM does not carry sufficient kinetic energy to trigger a transition across the band gap, which is $\sim6$\,eV in BN.

\section{Single Phonon Excitations}
\label{sec:phonon}

Finally, we derive single phonon production rates following the same procedure. Assuming zero temperature, the initial state is the ground state with no phonons, and the final state contains one phonon:
\be
|i\rangle = |0\rangle\,,\qquad
|f\rangle = |\nu,\vect{k}\rangle = \hat a_{\nu,\vect{k}}^\dagger |0\rangle\,,
\ee
where the canonical commutation relations read $\bigl[\hat a_{\nu,\vect{k}},\, \hat a_{\nu',\vect{k}'}^\dagger\bigr]=\delta_{\nu\nu'}\delta_{\vect{k}\vect{k}'}$, etc. Note that phonons are labeled by a branch index $\nu = 1,\dots, 3n$, where $n$ is the number of atoms/ions in each primitive cell, and a momentum vector $\vect{k}$ within the first Brillouin zone. For a crystal with $N$ primitive unit cells, $\vect{k}$ takes $N$ discrete values. In the end we take the limit $N\to\infty$, where $\vect{k}$ becomes continuous.

To see how ${\cal F}_T(\vect{q})$ should be quantized in the phonon Hilbert space, we note that phonons arise from atom/ion displacements:
\be
\vect{u}_{lj} = \vect{x}_{lj}-\vect{x}_{lj}^0 = \sum_\nu \sum_{\vect{k}\in\text{1BZ}} \frac{1}{\sqrt{2Nm_j\omega_{\nu,\vect{k}}}} \Bigl( \,\hat a_{\nu,\vect{k}}\,\vect{\epsilon}_{\nu,\vect{k},j} \,e^{i\vect{k}\cdot\vect{x}_{lj}^0} +\hat a^\dagger_{\nu,\vect{k}}\, \vect{\epsilon}_{\nu,\vect{k},j}^* \,e^{-i\vect{k}\cdot\vect{x}_{lj}^0} \Bigr)\,,
\label{eq:ulj}
\ee
where $\vect{x}_{lj}$ is the position of the $j$th atom/ion in the $l$th primitive cell, $\vect{x}_{lj}^0$ is the equilibrium position, $m_j$ are the atom/ion masses, $\omega_{\nu,\vect{k}}$ are the phonon energies, and $\vect{\epsilon}_{\nu,\vect{k},j}$ are the phonon polarization vectors, normalized such that $\sum_j|\vect{\epsilon}_{\nu,\vect{k},j}|^2=1$. The task is thus to find how ${\cal F}_T(\vect{q})$ depends on the atom/ion positions $\vect{x}_{lj}$ and displacements $\vect{u}_{lj}$.

To do so, let us revisit the scattering potential in Eq.~\eqref{eq:vx}. For a periodic crystal, it can be written as a sum over contributions from individual atoms/ions:
\beq
\mathcal{V}(\vect{x}) &=& \sum_{l,j} \int_{\Omega_{lj}} d^3x' \bigl[ n_p^{lj}(\vect{x}') \mathcal{V}_p(\vect{x}-\vect{x}') + n_n^{lj}(\vect{x}')\, \mathcal{V}_n(\vect{x}-\vect{x}') + n_e^{lj}(\vect{x}')\, \mathcal{V}_e(\vect{x}-\vect{x}') \bigr] \nonumber\\
&=& \sum_{l,j} \int_{\Omega_{lj}} d^3r \bigl[ n_p^{lj}(\vect{r}) \mathcal{V}_p(\vect{x}-\vect{x}_{lj}-\vect{r}) + n_n^{lj}(\vect{r})\, \mathcal{V}_n(\vect{x}-\vect{x}_{lj}-\vect{r}) + n_e^{lj}(\vect{r})\, \mathcal{V}_e(\vect{x}-\vect{x}_{lj}-\vect{r}) \bigr]\,,\nonumber\\
\label{eq:vx-phonon}
\eeq
where $\Omega_{lj}$ is a volume surrounding the lattice site $l,j$. Within each site volume, we have changed the integration variable to $\vect{r}=\vect{x}'-\vect{x}_{lj}$, the position relative to the center of the site, and defined $n_p^{lj}(\vect{r})\equiv n_p(\vect{x}_{lj}+\vect{r})$, etc. For protons and neutrons, $n_{p,n}^{lj}$ here coincides with $n_{p,n}^0$ introduced in Sec.~\ref{sec:nuclear} for the nucleus at site $l,j$. Also, displacing an atom/ion does not change the nucleon distributions inside of a nucleus. Thus, we can write
\be
n_{p,n}^{lj}(\vect{r})=n_{p,n}^j(\vect{r})\,,
\ee
which makes it clear that nucleon number densities are the same in all primitive cells, and are not affected by atom/ion displacements in any particular primitive cell. For electrons, on the other hand, this is generally not true, since electron wavefunctions are distorted when displacing an atom/ion relative to the other atoms/ions in the crystal lattice. To account for this effect, we write
\be
n_e^{lj}(\vect{r}) = n_e^j(\vect{r}) + \sum_{l',j'}\frac{\delta n_e^{lj}}{\delta\vect{u}_{l'j'}}\cdot \vect{u}_{l'j'} +\OO(u^2)
\simeq n_e^j(\vect{r}) +\frac{\delta n_e^{lj}(\vect{r})}{\delta\vect{u}_{lj}}\cdot \vect{u}_{lj}\,,
\label{eq:nelj}
\ee
where the last expression assumes the effect of electron redistribution following an atom/ion displacement is weak and local. This is usually a good approximation for ionic crystals such as gallium arsenide (GaAs), where electrons are semi-localized, and displacing an ion tends not to significantly affect the electron clouds of neighboring ions. For covalent crystals such as silicon, valence electron wavefunctions are more disperse, so more terms in the $l'j'$ sum should be included for an accurate calculation. 

Assuming the approximation in Eq.~\eqref{eq:nelj} is valid, we can Fourier transform Eq.~\eqref{eq:vx-phonon} and obtain
\beq
\widetilde{\cal V}(-\vect{q}) &=& \MM_0(q) \sum_{l,j} e^{i\vect{q}\cdot\vect{x}_{lj}} \biggl[ f_p \widetilde n_p^{j}(-\vect{q}) +f_n \widetilde n_n^{j}(-\vect{q}) +f_e \widetilde n_e^{j}(-\vect{q}) +f_e\,\frac{\delta \widetilde n_e^{lj}(-\vect{q})}{\delta\vect{u}_{lj}}\cdot \vect{u}_{lj} \biggr] \nonumber\\
&=& \MM_0(q) \sum_{l,j} e^{i\vect{q}\cdot\vect{x}_{lj}} \biggl[ f_j F_{N_j}(\vect{q}) +f_e \widetilde n_e^{j}(-\vect{q}) +f_e\,\frac{\delta \widetilde n_e^{lj}(-\vect{q})}{\delta\vect{u}_{lj}}\cdot \vect{u}_{lj} \biggr]\,,
\eeq
where $f_j = f_p Z_j +f_n(A_j-Z_j)$, and $F_{N_j}(\vect{q})$ is the nuclear form factor (introduced in Sec.~\ref{sec:nuclear}) for the nucleus occupying site $j$ in each primitive cell. We therefore obtain
\be
{\cal F}_T(\vect{q}) = \sum_{l,j} \bigl[{\cal F}_j^0(\vect{q}) + \vect{\Delta}_j(\vect{q}) \cdot \vect{u}_{lj}\bigr] \,e^{i\vect{q}\cdot\vect{x}_{lj}}\,,
\label{eq:ft-phonon}
\ee
with
\be
{\cal F}_j^0(\vect{q}) \equiv \frac{1}{f_\psi^0} \bigl[ f_j F_{N_j}(\vect{q}) +f_e \widetilde n_e^{j}(-\vect{q}) \bigr]\,,\qquad
\vect{\Delta}_j(\vect{q}) \equiv \frac{f_e}{f_\psi^0} \frac{\delta \widetilde n_e^{lj}(-\vect{q})}{\delta\vect{u}_{lj}}\,,
\label{eq:Fj0Deltaj}
\ee
where $f_\psi^0=f_n^0$ ($f_e^0$) if the rate is written in terms of $\overline{\sigma}_n$ ($\overline{\sigma}_e$). Note that $\vect{\Delta}_j$ is independent of $l$ due to lattice translation symmetries. From Eq.~\eqref{eq:ft-phonon} we see that ${\cal F}_T(\vect{q})$ depends on $\vect{u}_{lj}$ -- which are quantized in terms of phonon modes as in Eq.~\eqref{eq:ulj} -- via both the phase factor $e^{i\vect{q}\cdot\vect{x}_{lj}}=e^{i\vect{q}\cdot(\vect{x}_{lj}^0+\vect{u}_{lj})}$ and the $\vect{\Delta}_j(\vect{q}) \cdot \vect{u}_{lj}$ term. 

With ${\cal F}_T(\vect{q})$ quantized in the phonon Hilbert space, we now move on to calculate the matrix element $\langle \nu,\vect{k}|{\cal F}_T(\vect{q})|0\rangle$. We first apply the Baker-Campbell-Hausdorff (BCH) formula to the phase factor $e^{i\vect{q}\cdot\vect{x}_{lj}}$ to move annihilation operators to the right:
\beq
e^{i\vect{q}\cdot\vect{x}_{lj}} &=& e^{i\vect{q}\cdot\vect{x}_{lj}^0} \prod_{\nu,\vect{k}}
\,\exp \biggl[ \frac{i(\vect{q}\cdot\vect{\epsilon}_{\nu,\vect{k},j}^*)\,e^{-i\vect{k}\cdot\vect{x}_{lj}^0}}{\sqrt{2Nm_j\omega_{\nu,\vect{k}}}} \hat a^\dagger_{\nu,\vect{k}} 
+ \frac{i(\vect{q}\cdot\vect{\epsilon}_{\nu,\vect{k},j})\,e^{i\vect{k}\cdot\vect{x}_{lj}^0}}{\sqrt{2Nm_j\omega_{\nu,\vect{k}}}}  \,\hat a_{\nu,\vect{k}} \biggr] \nonumber\\
&=& e^{i\vect{q}\cdot\vect{x}_{lj}^0} \prod_{\nu,\vect{k}} 
\,\exp \biggl[ \frac{i(\vect{q}\cdot\vect{\epsilon}_{\nu,\vect{k},j}^*)\,e^{-i\vect{k}\cdot\vect{x}_{lj}^0}}{\sqrt{2Nm_j\omega_{\nu,\vect{k}}}} \hat a^\dagger_{\nu,\vect{k}} \biggr] 
\cdot \exp\biggl[ \frac{i(\vect{q}\cdot\vect{\epsilon}_{\nu,\vect{k},j})\,e^{i\vect{k}\cdot\vect{x}_{lj}^0}}{\sqrt{2Nm_j\omega_{\nu,\vect{k}}}}  \,\hat a_{\nu,\vect{k}} \biggr] 
\times \nonumber\\
&& \qquad\qquad\;\, \exp\biggl( \frac{|\vect{q}\cdot\vect{\epsilon}_{\nu,\vect{k},j}|^2}{4Nm_j\omega_{\nu,\vect{k}}} \bigl[\hat a_{\nu,\vect{k}}^\dagger ,\, \hat a_{\nu,\vect{k}} \bigr] \biggr) \nonumber\\
&=& e^{i\vect{q}\cdot\vect{x}_{lj}^0} \,e^{-W_j(\vect{q})} 
\,\exp \biggl[\sum_{\nu,\vect{k}} \frac{i(\vect{q}\cdot\vect{\epsilon}_{\nu,\vect{k},j}^*)\,e^{-i\vect{k}\cdot\vect{x}_{lj}^0}}{\sqrt{2Nm_j\omega_{\nu,\vect{k}}}} \hat a^\dagger_{\nu,\vect{k}} \biggr] 
\exp\biggl[ \sum_{\nu,\vect{k}}\frac{i(\vect{q}\cdot\vect{\epsilon}_{\nu,\vect{k},j})\,e^{i\vect{k}\cdot\vect{x}_{lj}^0}}{\sqrt{2Nm_j\omega_{\nu,\vect{k}}}}  \,\hat a_{\nu,\vect{k}} \biggr] \,,\quad
\label{eq:exp-expand}
\eeq
where we have used the fact that the commutator between creation and annihilation operators is a classical number so the BCH series terminates. In the last equation,
\be
W_j (\vect{q}) = \frac{1}{4Nm_j} \sum_\nu \sum_{\vect{k}\in\text{1BZ}} \frac{|\vect{q}\cdot\vect{\epsilon}_{\nu,\vect{k},j}|^2}{\omega_{\nu,\vect{k}}}
\rightarrow \frac{\Omega}{4m_j}\sum_\nu \int_\text{1BZ}\frac{d^3k}{(2\pi)^3} \frac{|\vect{q}\cdot\vect{\epsilon}_{\nu,\vect{k},j}|^2}{\omega_{\nu,\vect{k}}}
\label{eq:Wj}
\ee
is the Debye-Waller factor (in the continuum limit $\sum_{\vect{k}}\to V\int\frac{d^3k}{(2\pi)^3}=N\Omega\int\frac{d^3k}{(2\pi)^3}$ with $\Omega$ the volume of the primitive cell). The physical meaning of this factor is that a transition $|i\rangle\to|f\rangle$ can be accompanied by additional phonons' creation out of the vacuum followed by their annihilation, and all these processes are resummed into the exponential. The matrix element thus becomes
\beq
\langle \nu,\vect{k}|{\cal F}_T(\vect{q})|0\rangle &=& \sum_{l,j}e^{i\vect{q}\cdot\vect{x}_{lj}^0} \,e^{-W_j(\vect{q})}\times \nonumber\\[-8pt]
&&\qquad \langle \nu,\vect{k}| \bigl[{\cal F}_j^0(\vect{q}) + \vect{\Delta}_j(\vect{q}) \cdot \vect{u}_{lj}\bigr] \exp \biggl[\sum_{\nu',\vect{k}'} \frac{i(\vect{q}\cdot\vect{\epsilon}_{\nu',\vect{k}',j}^*)\,e^{-i\vect{k}'\cdot\vect{x}_{lj}^0}}{\sqrt{2Nm_j\omega_{\nu',\vect{k}'}}} \hat a^\dagger_{\nu',\vect{k}'} \biggr] |0\rangle \nonumber\\
&=& \sum_{l,j} e^{i(\vect{q}-\vect{k})\cdot\vect{x}_{lj}^0} \,e^{-W_j(\vect{q})} \frac{i}{\sqrt{2Nm_j\omega_{\nu,\vect{k}}}} \times \nonumber\\
&&\qquad \biggl[ {\cal F}_j^0 \vect{q} -i\,\vect{\Delta}_j
+\frac{\vect{q}}{Nm_j} \sum_{\nu',\vect{k}'}\frac{(i\vect{\Delta}_j\cdot\vect{\epsilon}_{\nu',\vect{k}',j})(\vect{q}\cdot\vect{\epsilon}_{\nu',\vect{k}',j}^*)}{2\omega_{\nu',\vect{k}'}} 
\biggr] \cdot\vect{\epsilon}_{\nu,\vect{k},j}^*\,.
\eeq
The $l$ sum can be eliminated via the identity
\be
\sum_l e^{i(\vect{q}-\vect{k})\cdot\vect{x}_l}=N\sum_{\vect{G}} \delta_{\vect{q}-\vect{k},\vect{G}}\,,
\label{eq:lsum}
\ee
where $\vect{x}_{lj}^0=\vect{x}_l+\vect{x}_j^0$ with $\vect{x}_l$ being the position of the $l$th primitive cell and $\vect{x}_j^0$ being the equilibrium position of the $j$th atom/ion within the primitive cell, and $\vect{G}$ runs over the reciprocal lattice vectors. In fact, at most one term in the $\vect{G}$ sum is picked out for given $\vect{q}$ and $\vect{k}$, since $\vect{k}\in$ 1BZ. We will thus drop the $\vect{G}$ sum in what follows. On each phonon branch, as we sum over $\vect{k}$, only the mode that satisfies $\vect{q}=\vect{k}+\vect{G}$ can give a nonzero contribution to the dynamic structure factor, as a result of lattice momentum conservation.

It is worth emphasizing that the notion of momentum conservation here differs from the one familiar in particle physics, due to the spontaneous breaking of continuous translation symmetries. While each phonon can be thought of as carrying a momentum $\vect{k}$ within the 1BZ, it can be excited even when the momentum transfer $\vect{q}$ is outside the 1BZ via Umklapp scattering, in which case $\vect{G}\ne\vect{0}$. For DM heavier than $\sim$ MeV, the momentum transfer can exceed $\sim$ keV, the typical size of the 1BZ. In this case, Umklapp processes can contribute significantly if the matrix element has support at high $q$ (which is the case for a heavy mediator). We will see an example of this in Sec.~\ref{sec:phonon-n}. Note that momentum is still conserved at the fundamental level: the extra momentum $\vect{G}$ leads to a recoil of the entire crystal, which becomes unobservable in the limit $N\to\infty$. On the other hand, the notion of energy conservation is the same, as continuous time translation symmetry remains unbroken. As a result, the energy deposition has to match the phonon energy for a phonon mode to be excited.

With the equations above, we obtain the dynamic structure factor:
\beq
S(\vect{q},\omega) &=& \frac{2 \pi}{V} \sum_\nu \sum_{\vect{k}\in\text{1BZ}} \bigl| \langle \nu,\vect{k}| {\cal F}_T(\vect{q})| 0\rangle \bigr|^2 \,\delta\bigl(\omega-\omega_{\nu,\vect{k}}\bigr) \nonumber\\
&=& \frac{\pi}{\Omega} \sum_\nu \frac{1}{\omega_{\nu,\vect{k}}} \biggl| \sum_j \frac{e^{-W_j(\vect{q})}}{\sqrt{m_j}}\,e^{i\vect{G}\cdot \vect{x}_j^0} \bigl(\vect{Y}_j\cdot\vect{\epsilon}_{\nu,\vect{k},j}^*\bigr)\biggr|^2 \,\delta\bigl(\omega-\omega_{\nu,\vect{k}}\bigr)\,,
\label{eq:sfun-phonon}
\eeq
where
\be
\vect{Y}_j \equiv {\cal F}_j^0 \vect{q} -i\,\vect{\Delta}_j
+\frac{\Omega}{m_j} \,\vect{q}\, \sum_{\nu'}\int_\text{1BZ}\frac{d^3k'}{(2\pi)^3}\,\frac{(i\vect{\Delta}_j\cdot\vect{\epsilon}_{\nu',\vect{k}',j})(\vect{q}\cdot\vect{\epsilon}_{\nu',\vect{k}',j}^*)}{2\omega_{\nu',\vect{k}'}} \,.
\label{eq:Yj}
\ee
We have made it implicit in the last line of Eq.~\eqref{eq:sfun-phonon} that the $\vect{k}$ vector is the one inside the first Brillouin zone that satisfies $\vect{q}=\vect{k}+\vect{G}$.

Finally, integrating over the DM velocity distribution, we obtain the rate per target mass:
\be
R = \frac{1}{m_\text{cell}} \frac{\rho_\chi}{m_\chi} \frac{\pi\overline\sigma}{2\mu^2} \int\frac{d^3q}{(2\pi)^3} \,{\cal F}_\text{med}^2(q) \sum_\nu \frac{1}{\omega_{\nu,\vect{k}}}\, \biggl|\sum_{j} \frac{e^{-W_j(\vect{q})}}{\sqrt{m_j}} \,e^{i\vect{G}\cdot\vect{x}_j^0} \bigl(\vect{Y}_j\cdot\vect{\epsilon}_{\nu,\vect{k},j}^*\bigr) \biggr|^2 \,g(\vect{q},\omega_{\nu,\vect{k}}) \,,
\label{eq:Rphonon}
\ee
where $m_\text{cell}=\rho_T\Omega$ is the mass contained in a primitive cell. The mediator form factor ${\cal F}_\text{med}$, the Debye-Waller factor $W_j(\vect{q})$ and the $g(\vect{q},\omega)$ function are given by Eqs.~\eqref{eq:Fmed}, \eqref{eq:Wj} and \eqref{eq:gfun}, respectively. The DM couplings are encoded in the $\vect{Y}_j$ vectors given in Eq.~\eqref{eq:Yj}, with ${\cal F}_j^0,\vect{\Delta}_j$ defined in Eq.~\eqref{eq:Fj0Deltaj}. Meanwhile, the material specific quantities -- phonon dispersions $\omega_{\nu,\vect{k}}$ and polarization vectors $\vect{\epsilon}_{\nu,\vect{k},j}$ -- can be numerically computed using DFT methods detailed in our companion paper~\cite{Griffin:2019mvc}.

In the following subsections, we discuss the phonon excitation calculation in more detail. It is clear from the discussion above that $\vect{Y}_j$ are the key quantities to compute for any specific DM model. In Sec.~\ref{sec:phonon-n}, we consider the simpler case where DM couples only to nucleons but not electrons, and point out an interesting complementarity with nuclear recoils. We also discuss the relevance of Umklapp processes for DM heavier than an MeV, for both heavy and light mediators. Including DM-electron couplings introduces complications, but we show in Sec.~\ref{sec:phonon-e} that $\vect{Y}_j$ take a simple form in the low $q$ limit for general couplings $f_{p,n,e}$. Note that the dark photon mediator benchmark ($f_p^0=f_e^0, f_n^0=0$) has been studied in Refs.~\cite{Knapen:2017ekk,Griffin:2018bjn} based on the Fr\"ohlich Hamiltonian. Our calculation here reproduces previous results, and helps clarify their range of validity.

\subsection{Dark Matter Coupling Only to Nucleons}
\label{sec:phonon-n}

Setting $f_e=0$ and $f_\psi^0=f_n^0=f_n$ in Eq.~\eqref{eq:Fj0Deltaj}, we have
\be
{\cal F}_j^0(\vect{q}) = \biggl( \frac{f_j}{f_n} \biggr) \,F_{N_j}(q)\,,\qquad
\vect{\Delta}_j(\vect{q})=\vect{0}\,.
\ee
In this case, $\vect{Y}_j$ is simply ${\cal F}_j^0\vect{q}$, and the rate Eq.~\eqref{eq:Rphonon} becomes
\beq
R &=& \frac{1}{m_\text{cell}} \frac{\rho_\chi}{m_\chi} \frac{\pi\overline\sigma_n}{2\mu_{\chi n}^2} \int\frac{d^3q}{(2\pi)^3} \, {\cal F}_\text{med}^2(q) \times \nonumber\\
&&\qquad\qquad \sum_\nu \frac{1}{\omega_{\nu,\vect{k}}}\, \biggl|\sum_{j} \frac{e^{-W_j(\vect{q})}}{\sqrt{m_j}}\frac{f_j}{f_n} \,F_{N_j}(q) \,e^{i\vect{G}\cdot\vect{x}_j^0} \bigl(\vect{q}\cdot\vect{\epsilon}_{\nu,\vect{k},j}^*\bigr) \biggr|^2 \,g(\vect{q},\omega_{\nu,\vect{k}})\,.
\label{eq:phonon-n-rate}
\eeq

It is interesting to compare to the nuclear recoils case. If there is only one atom in the primitive cell, we have $m_\text{cell}=m_j=m_N$, and
\be
R = \frac{\rho_\chi}{m_\chi} \frac{\pi\overline\sigma_n}{2\mu_{\chi n}^2} \int\frac{d^3q}{(2\pi)^3} \,e^{-2W} \,\frac{f_N^2}{f_n^2} F_{N}^2 {\cal F}_\text{med}^2 \sum_\nu  \frac{\bigl|\vect{q}\cdot\vect{\epsilon}_{\nu,\vect{k}}^*\bigr|^2}{m_N^2\omega_{\nu,\vect{k}}} \, g(\vect{q},\omega_{\nu,\vect{k}})\,.
\ee
The differential rate reads
\be
\frac{dR}{d\omega} = \frac{\rho_\chi}{m_\chi} \frac{\pi\overline\sigma_n}{2\mu_{\chi n}^2} \int\frac{d^3q}{(2\pi)^3} \,e^{-2W}\,\frac{f_N^2}{f_n^2} F_{N}^2 {\cal F}_\text{med}^2 \, g(\vect{q},\omega)\sum_\nu  \frac{\bigl|\vect{q}\cdot\vect{\epsilon}_{\nu,\vect{k}}^*\bigr|^2}{m_N^2\omega} \,\delta(\omega-\omega_{\nu,\vect{k}})\,.\quad
\label{eq:phonon-n-monatomic}
\ee
On the other hand, we can rewrite Eq.~\eqref{eq:dRdED-nr-1} for nuclear recoils in terms of the $g(\vect{q}, \omega)$ function via $\int q\, dq\,\eta(v_\text{min}) \to 2\int\frac{d^3q}{(2\pi)^3}\,g(\vect{q},\omega)$, and multiply the integrand by $1=\frac{q^2}{2m_N\omega}$:
\be
\frac{dR}{d\omega} = \frac{\rho_\chi}{m_\chi} \frac{\pi\overline\sigma_n}{2\mu_{\chi n}^2} \int\frac{d^3q}{(2\pi)^3}\,\frac{f_N^2}{f_n^2} F_{N}^2 {\cal F}_\text{med}^2\, g(\vect{q},\omega) \frac{q^2}{m_N^2\omega} \,\delta\biggl(\omega-\frac{q^2}{2m_N}\biggr) \qquad\text{(nuclear recoil)}\,.
\label{eq:phonon-n-nuclearrecoil}
\ee
One can clearly see the similarity between Eqs.~\eqref{eq:phonon-n-monatomic} and \eqref{eq:phonon-n-nuclearrecoil}. However, a key difference between nuclear recoils and phonon excitations is the way in which contributions from different atoms add up in the case of more than one atoms in the primitive cell. Comparing Eq.~\eqref{eq:phonon-n-rate} against Eq.~\eqref{eq:dRdED-nr-2}, we see that, in contrast to the nuclear recoils case where we add up the {\it rates} from inequivalent nuclei, for phonon excitations the sum over $j$ is taken at the {\it amplitude} level. It is worth noting, however, that this apparent coherence does not result in a more favorable scaling of the detection rate. In fact, the total rate per target mass scales with neither the number of nuclei in the primitive cell, nor the total number of atoms/ions in the crystal. The former can be seen from the fact that phonon polarization vectors scale as $\vect{\epsilon}_{\nu,\vect{k},j}\sim \sqrt{m_j/m_\text{cell}}$, which, together with the prefactor, means the denominator of Eq.~\eqref{eq:phonon-n-rate} scales as $m_\text{cell}^2$. The latter is because of the $1/\sqrt{N}$ normalization factor when expanding $\vect{u}_{lj}$ in terms of phonon creation and annihilation operators (see Eq.~\eqref{eq:ulj}). The intuition here is that, despite the collective nature of phonon excitations, we have to project the motion of each atom onto the phonon modes that match the energy-momentum transfer. As a result, coherence between more atoms comes with a price of a smaller overlap with phonon modes.

Another key difference between nuclear recoils and phonon excitations, alluded to in Fig.~\ref{fig:kin} and Sec.~\ref{sec:nuclear-validity}, is the kinematic regimes probed. In the phonon case, the Debye-Waller factor $e^{-W_j}$ cuts off the momentum integral for $q \gtrsim \sqrt{m_N\omega_\text{ph}}$, the inverse spatial extent of the nucleus wavefunction. The $\omega_\text{ph}$ here should be thought of as an average phonon energy over the entire 1BZ, which is of the same order as $\omega_\text{ph}^\text{max}$. As discussed in Sec.~\ref{sec:nuclear-validity}, this high $q$ regime is exactly where the nuclear recoil calculation becomes valid. In addition, nuclear recoils happen at much higher energy depositions $\omega =q^2/2m_N\gg \omega_\text{ph}^\text{max}$ than phonon excitations.

\begin{figure}[t]
\includegraphics[width=0.48\linewidth]{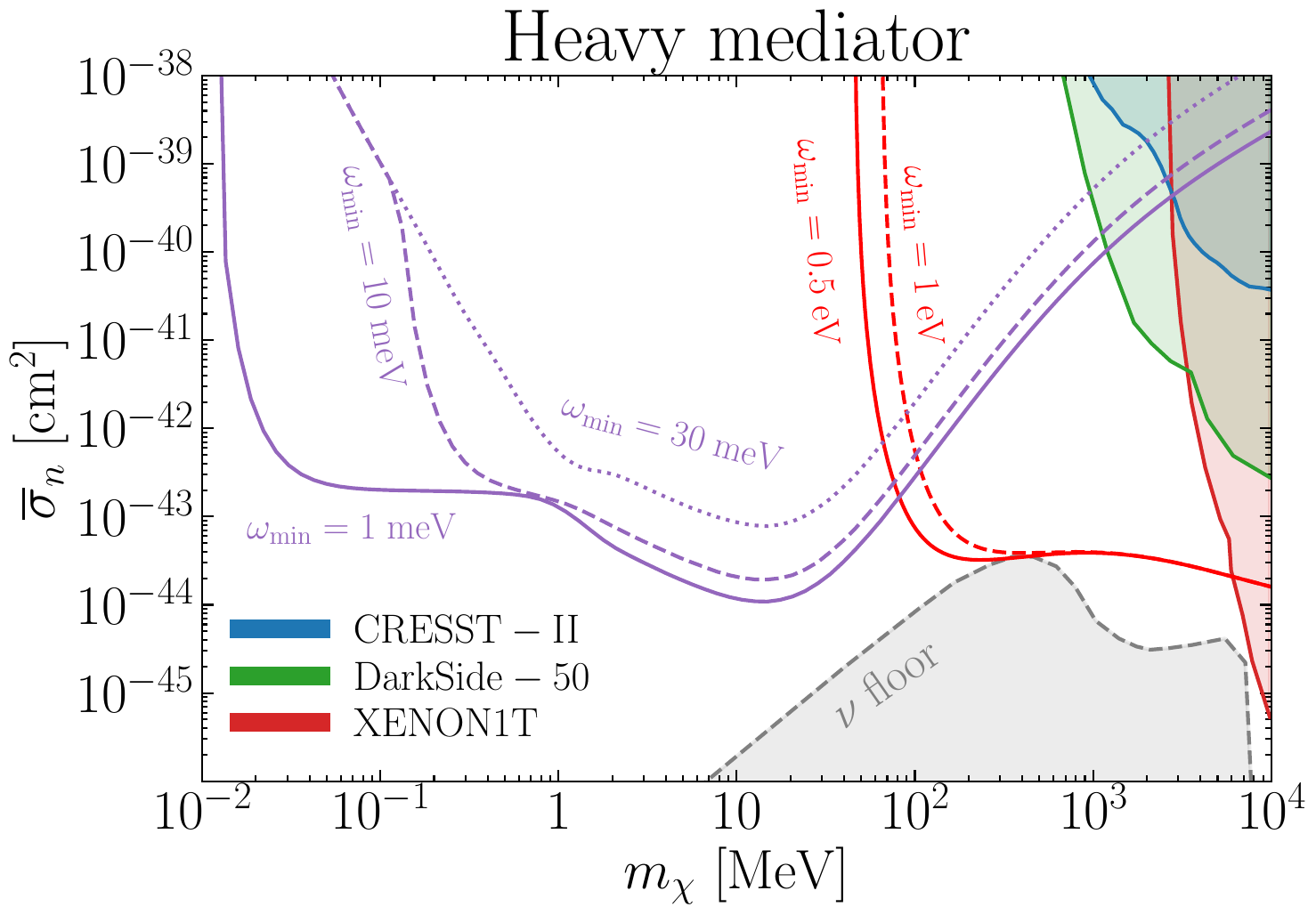}\hspace{0.03\linewidth}
\includegraphics[width=0.48\linewidth]{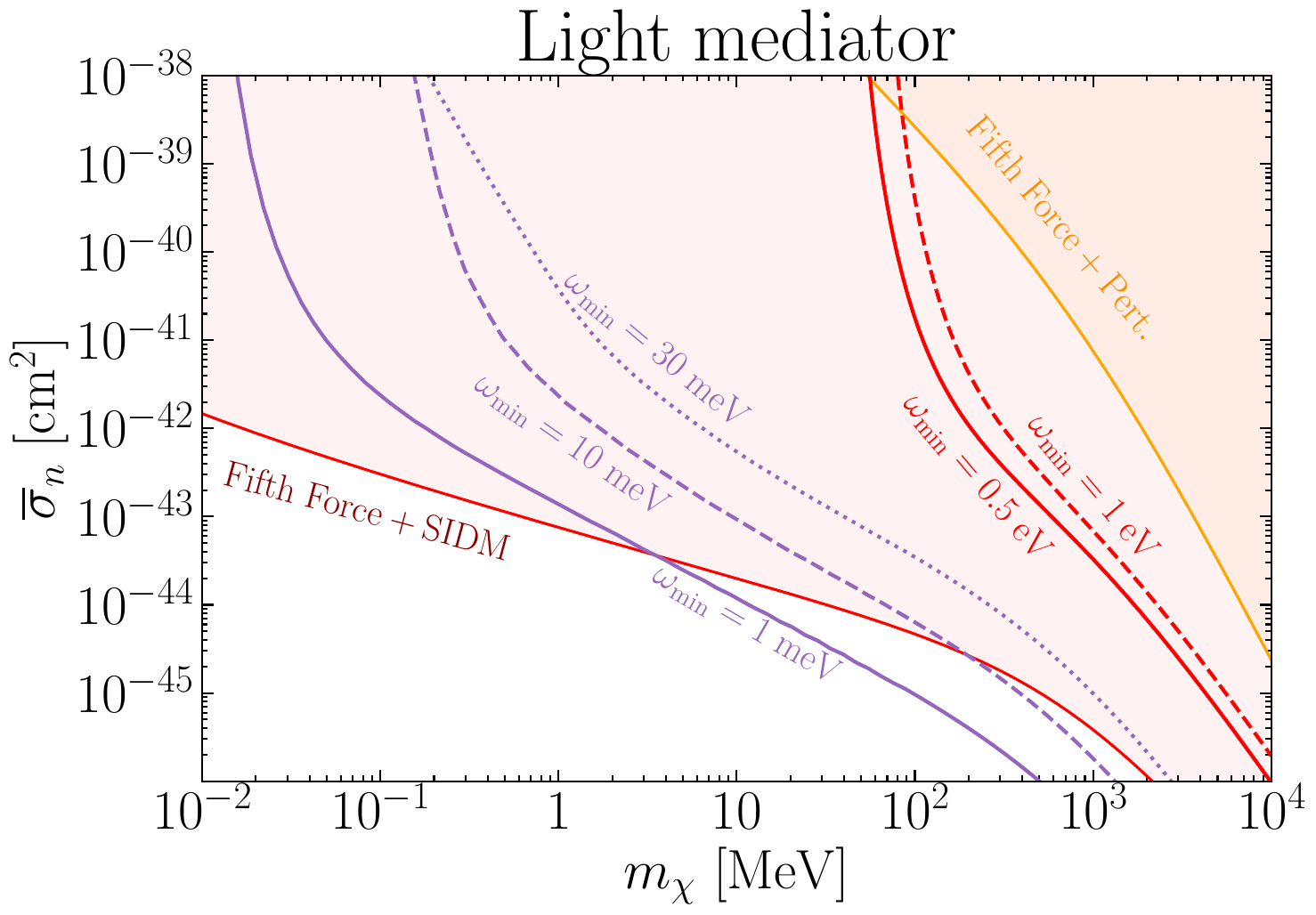}
\caption{\label{fig:sigman}
    Projected reach for DM scattering via a heavy (left, $m_\phi \gtrsim 400$ MeV) or light (right, $m_\phi = 1$ eV) scalar mediator coupling to nucleons ($f_p=f_n$, $f_e=0$), assuming 1\,kg-yr exposure with a GaAs target, 3 signal events and no background. Both single phonon production (purple, assuming energy thresholds $\omega_\text{min}=1,10,30\,\text{meV}$) and nuclear recoils (red, assuming $\omega_\text{min}=0.5,1\,\text{eV}$) are complementary in probing currently unconstrained parameter space. The heavy mediator case is free from stellar constraints for $m_\phi \gtrsim 400$ MeV~\cite{Knapen:2017xzo}, and the neutrino floor is taken from Ref.~\cite{Battaglieri:2017aum}. Currently, the best experimental nuclear recoil constraints in this region of parameter space are from DarkSide-50~\cite{Agnes2018b} (assuming binomial fluctuations), and XENON1T (combined limits from~\cite{Aprile2018a, Aprile:2019xxb}). We also show the constraint from CRESST-II~\cite{Angloher:2015ewa}, which is stronger than the DarkSide-50 constraint at low masses assuming no fluctuation in energy quenching. A more complete collection of nuclear recoil constraints can be found in Refs.~\cite{Agnes2018b,Akerib2019b,Aprile:2019xxb}. For a light mediator with $m_\phi = 1$ eV, fifth force experiments provide the dominant constraint on mediator-nucleon couplings~\cite{Knapen:2017xzo}. Meanwhile, the mediator-$\chi$ coupling is constrained by DM self interactions (SIDM) if $\chi$ makes up all the DM~\cite{Knapen:2017xzo}, or just by perturbativity (Pert.)\ if $\chi$ is a DM subcomponent (in which case the projected reach can be easily rescaled).\footref{foot:one}
}
\end{figure}

A multi-channel search can exploit this complementarity between nuclear recoils and phonon excitations. Let us consider, as a benchmark model, a hadrophilic scalar mediator coupling identically to protons and neutrons ($f_p=f_n$, $f_e=0$). In Fig.~\ref{fig:sigman}, we compare the reach of the two channels, using GaAs as an example target material.\footnote{\label{foot:one}As discussed in arXiv v3 of the companion paper~\cite{Griffin:2019mvc}, changes in the constraint projections via single phonon excitations in Fig.~\ref{fig:sigman}, relative to previous versions, are due to a bug fix in computing $F_{N_j}$ which altered the constraints by a factor of, approximately, 2.25. The calculations have been updated using \textsf{PhonoDark} v1.1.0~\cite{phonodark}.} For a heavy mediator (left panel), we see that with sub-eV energy thresholds, nuclear recoils can probe DM masses above $\sim100\,$MeV --- this is the mass regime where the single phonon excitation rate suffers from Debye-Waller suppression. Below $\sim100\,$MeV where nuclear recoils lose sensitivity, single phonon excitations can probe a few more orders of magnitude of $m_\chi$, depending on the energy threshold. For a light mediator (right panel), on the other hand, single phonon excitations outperform nuclear recoils for all $m_\chi$. This is because the momentum integral is dominated by the lowest $q$, which only depends on the energy threshold, $q_\text{min} \simeq \omega_\text{min}/v_\text{max}$. The mass scaling of the curves in Fig.~\ref{fig:sigman} can be understood with a close examination of phase space integrals; we reserve a detailed discussion, including how the various features of the curves depend on material properties, for the companion paper~\cite{Griffin:2019mvc}.

\begin{figure}[t]
\includegraphics[width=\linewidth]{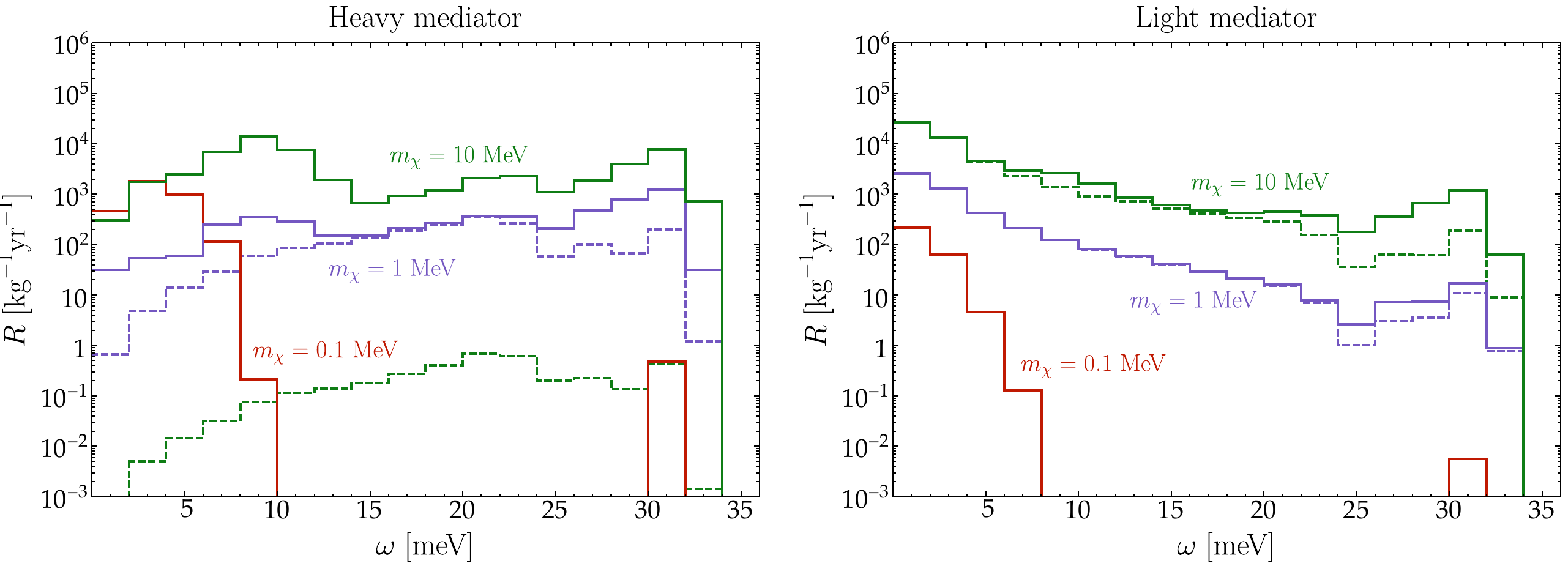}
\caption{\label{fig:umklapp}
Differential rate of single phonon excitations in a GaAs target for $m_\chi=0.1, 1, 10\,$MeV, assuming a heavy (left) or light (right) scalar mediator coupling to nucleons ($f_p=f_n$, $f_e=0$), with $\bar\sigma_n=10^{-40}\text{ cm}^2$ and $\omega_\text{min}=1\,$meV. Contributions from momentum transfer within the first Brillouin zone are shown in dash. Umklapp processes account for the differences between solid and dashed histograms.
}
\end{figure}

It is also worth noting that while direct production of single phonons has been proposed mainly as a channel to search for sub-MeV DM, we see from Fig.~\ref{fig:sigman} that its sensitivity extends well above MeV DM masses, which is important for covering the parameter space out of reach in nuclear recoils. A DM particle heavier than $\sim$ MeV carries a momentum larger than the typical size of the 1BZ (or equivalently, the inverse lattice spacing). However, as explained below Eq.~\eqref{eq:lsum}, a crystal target is able to absorb a momentum transfer beyond the 1BZ while still producing a phonon, provided the energy deposition matches that of the phonon energy. Such Umklapp processes can contribute significantly to the rate. In Fig.~\ref{fig:umklapp}, we examine the role of Umklapp scattering by comparing the full rate (solid) vs.\ contributions from $q\in$ 1BZ (dashed), for three DM masses. We show the differential distribution up to 34\,meV, the highest phonon energy in GaAs. For $m_\chi=0.1\,$MeV, the maximum momentum transfer $q_\text{max} \simeq 2m_\chi v_\text{max}\simeq 0.56\,\text{keV}$ is within the 1BZ, so the solid and dashed histograms coincide. Also, only acoustic phonons with energies below $c_s q_\text{max}\simeq 9\,$meV (where $c_s$ is the speed of sound) and optical phonons are kinematically accessible; contributions from optical phonons are suppressed at low $q$~\cite{Cox:2019cod}, so the total rate is dominated by the low energy acoustic phonons. For $m_\chi=1\,$MeV and 10\,MeV, Umklapp processes dominate the rate in the heavy mediator case, since the momentum integral is dominated by large $q$. In the light mediator case, the matrix element peaks at small $q$, so the total rate is well approximated by the 1BZ contribution for sufficiently low energy thresholds (e.g.\ 1\,meV). However, Umklapp scattering can still contribute significantly in the highest energy bins, and dominate the rate if the energy threshold is higher (e.g.\ 30\,meV).

\subsection{Dark Matter With Couplings to Electrons}
\label{sec:phonon-e}

In the presence of electron couplings $f_e\ne 0$, information about electron distributions is needed for the rate calculation. We focus on ionic crystals in this subsection, for which Eq.~\eqref{eq:nelj} is a good approximation, and the rate formula Eq.~\eqref{eq:Rphonon} directly applies. In this case, we need $\widetilde n_e^j$ and $\delta\widetilde n_e^{lj}/\delta\vect{u}_{lj}$ as input. While $\widetilde n_e^j$ can be derived from the same electron wavefunctions as those used in electron transition calculations in Sec.~\ref{sec:electron}, $\delta\widetilde n_e^{lj}/\delta\vect{u}_{lj}$ is challenging to compute numerically for general $\vect{q}$ and $\vect{u}_{lj}$. 

However, the calculation simplifies in the limit $q\ll r_\text{ion}^{-1}$, the inverse ionic radii. As in classical electromagnetism, we can make a multipole expansion,
\be
\widetilde n_e^j (-\vect{q}) = \int_{\Omega_{lj}} d^3r\,e^{i\vect{q}\cdot\vect{r}} n_e^{lj}(\vect{r}) = N_{e,j} -i\vect{q}\cdot\vect{P}_{e,j} +\OO(q^2)\,,
\label{eq:netj}
\ee
where $N_{e,j}$ is the number of electrons associated with site $l,j$, and $\vect{P}_{e,j}$ is the electron contribution to the polarization in the volume $\Omega_{lj}$. Consider the response of the total polarization of the volume to a lattice displacement $\vect{u}_{lj}$:
\be
\delta\vect{P}_{lj} = Q_j\,\delta\vect{u}_{lj} +\delta\vect{P}_{e,j}\,,
\ee
where $Q_j = Z_j-N_{e,j}$ is the total charge. This defines the Born effective charge tensor:\footnote{More precisely, the Born effective charge $\vect{Z}^*_j$ is defined as the change in macroscopic polarization caused by a uniform displacement of the entire sublattice $j$~\cite{Wang/Vanderbilt:2007}. However, under the assumption we have made in Eq.~\eqref{eq:nelj} -- that the electrons respond locally to the ionic displacements -- the precise definition is equivalent to Eq.~\eqref{eq:Zjstar}.}
\be
\vect{Z}_j^* \equiv \frac{\delta \vect{P}_{lj}}{\delta\vect{u}_{lj}} = Q_j \mathbbm{1} +\frac{\delta \vect{P}_{e,j}}{\delta\vect{u}_{lj}}\,.
\label{eq:Zjstar}
\ee
Thus,
\be
\frac{\delta\widetilde n_e^{lj}(-\vect{q})}{\delta\vect{u}_{lj}} = -i\vect{q}\cdot\frac{\delta \vect{P}_{e,j}}{\delta\vect{u}_{lj}} +\OO(q^2) = -i\vect{q}\cdot(\vect{Z}_j^* - Q_j \mathbbm{1}) +\OO(q^2)\,.
\label{eq:dnedu}
\ee
From Eqs.~\eqref{eq:netj} and \eqref{eq:dnedu}, we obtain (choosing $f^0=f_e^0$ in the normalization):
\beq
{\cal F}_j^0(\vect{q}) &=& \frac{f_p}{f_e^0} Z_j +\frac{f_n}{f_e^0}(A_j-Z_j) +\frac{f_e}{f_e^0} N_{e,j} +\OO(q)\,,\\
\vect{\Delta}_j(\vect{q}) &=& -\frac{f_e}{f_e^0}\, i\vect{q}\cdot(\vect{Z}_j^* - Q_j \mathbbm{1}) +\OO(q^2)\,,
\eeq
where we have set $F_{N,j}(\vect{q})=1$ since $q\ll r_\text{ion}^{-1}$ is much smaller than the inverse nucleus radius. We therefore obtain the following simple expression for $\vect{Y}_j$:
\be
\vect{Y}_j = \vect{q}\cdot\biggl[\frac{f_p}{f_e^0} Z_j\, \mathbbm{1} +\frac{f_n}{f_e^0}(A_j-Z_j)\, \mathbbm{1} +\frac{f_e}{f_e^0} (Z_j \mathbbm{1} -\vect{Z}_j^*)\biggr] +\OO(q^2)\,.
\label{eq:Yj-lowq}
\ee

In the case of a vector or scalar mediator, the coupling ratios appearing in Eq.~\eqref{eq:Yj-lowq} should incorporate in-medium screening effects according to Eq.~\eqref{eq:screeningfactors}. As mentioned at the beginning of Sec.~\ref{sec:general-medium}, while dielectric response of an ionic crystal comes from both electrons and ions at phonon frequencies, only the electron contribution is included in the derivation of Eq.~\eqref{eq:screeningfactors}. That this is the correct treatment should be clear from the calculation above. Polarization induced by lattice displacements has been treated as an effective charge density $\nabla\cdot\vect{P}$, since it can induce the transition $|0\rangle\to|\nu,\vect{k}\rangle$. As such, it enters the source term rather than the dielectric matrix $\vect{\varepsilon}$ in Maxwell's equations. 
In the low $q$ limit, electron contributions to $\vect{\varepsilon}$ below the electronic band gap approach a constant $\vect{\varepsilon}_\infty$, referred to as the high-frequency dielectric constant.

In the special case of a dark photon mediator that kinetically mixes with the SM photon, $f_p^0=-f_e^0$, $f_n^0=0$. Combining Eqs.~\eqref{eq:Yj-lowq} and \eqref{eq:screeningfactors}, and setting $\vect{\varepsilon}\to\vect{\varepsilon}_\infty$, we obtain
\be
\vect{Y}_j = -\frac{q^2}{\vect{q}\cdot\vect{\varepsilon}_\infty\cdot\vect{q}}(\vect{q}\cdot\vect{Z}_j^*)\,.
\ee
By Eq.~\eqref{eq:Rphonon}, the rate is therefore
\beq
R &=& \frac{1}{m_\text{cell}} \frac{\rho_\chi}{m_\chi} \frac{\pi\overline\sigma_e}{2\mu_{\chi e}^2} \int\frac{d^3q}{(2\pi)^3} \,{\cal F}_\text{med}^2(q) \,\frac{q^4}{(\vect{q}\cdot\vect{\varepsilon}_\infty\cdot\vect{q})^2} \times\nonumber\\
&&\qquad\qquad \sum_\nu \frac{1}{\omega_{\nu,\vect{k}}}\, \biggl|\sum_{j} \frac{e^{-W_j(\vect{q})}}{\sqrt{m_j}} \,e^{i\vect{G}\cdot\vect{x}_j^0} \bigl(\vect{q}\cdot\vect{Z}_j^*\cdot\vect{\epsilon}_{\nu,\vect{k},j}^*\bigr) \biggr|^2 \,g(\vect{q},\omega_{\nu,\vect{k}}) \,.
\label{eq:Rphonon-dp}
\eeq
Note that since Eq.~\eqref{eq:Yj-lowq} for $\vect{Y}_j$ is derived in the limit $q\ll r_\text{ion}^{-1}\sim\OO(\text{keV})$, Eq.~\eqref{eq:Rphonon-dp} holds only when the integral is dominated by this region. This is the case for a light dark photon mediator for any DM mass, since the integrand peaks at small $q$. In this case, Eq.~\eqref{eq:Rphonon-dp} is in agreement with the result obtain in Ref.~\cite{Griffin:2018bjn} based on the Fr\"ohlich Hamiltonian. For a heavy mediator, on the other hand, the integrand peaks at $q_\text{max}=2m_\chi v_\text{max}$, so Eq.~\eqref{eq:Rphonon-dp} holds only for $m_\chi\ll(2v_\text{max}r_\text{ion})^{-1}\sim\OO(\text{MeV})$.

\begin{figure}[t]
\includegraphics[width=0.48\linewidth]{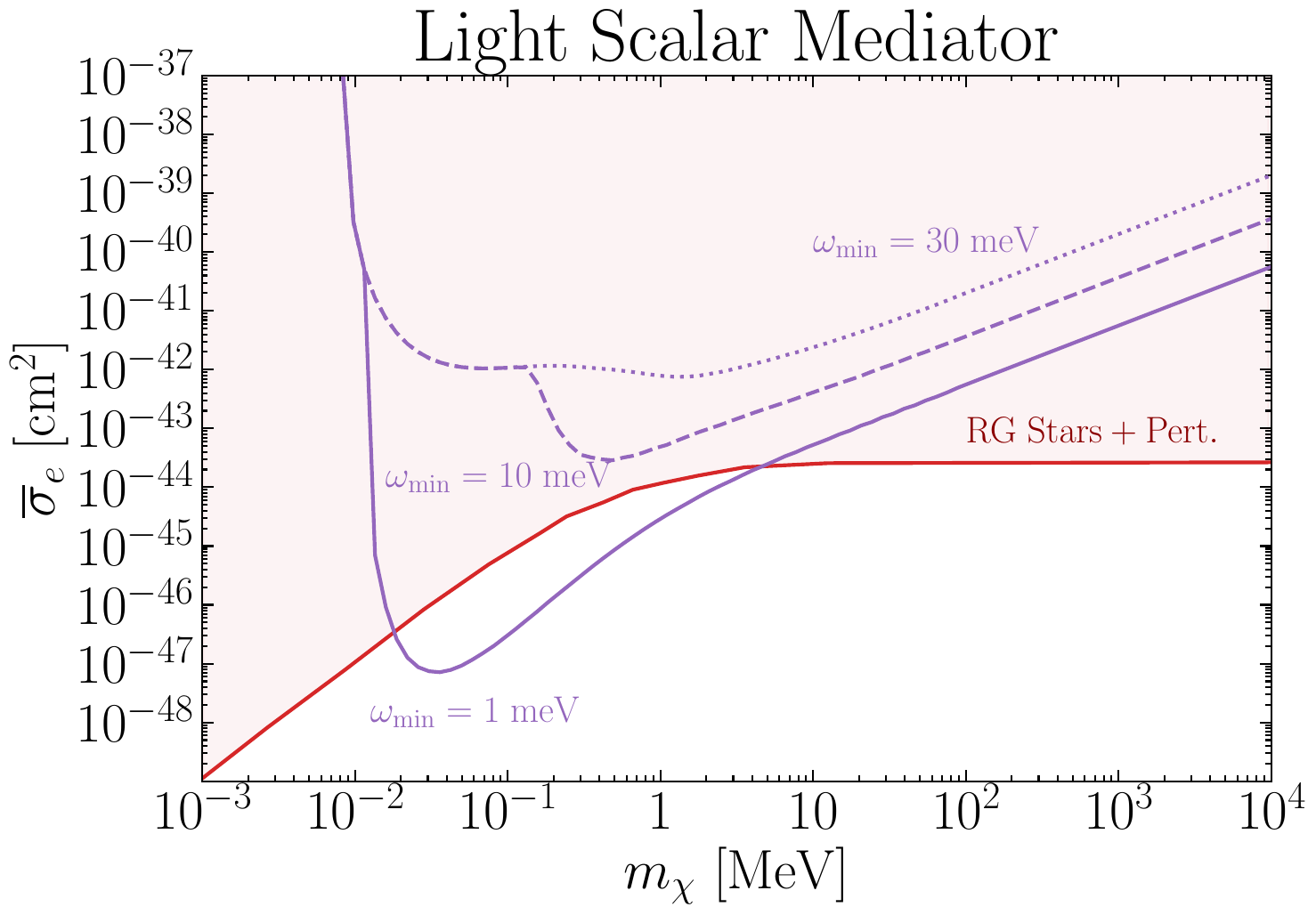}
\hspace{0.02\linewidth}
\includegraphics[width=0.48\linewidth]{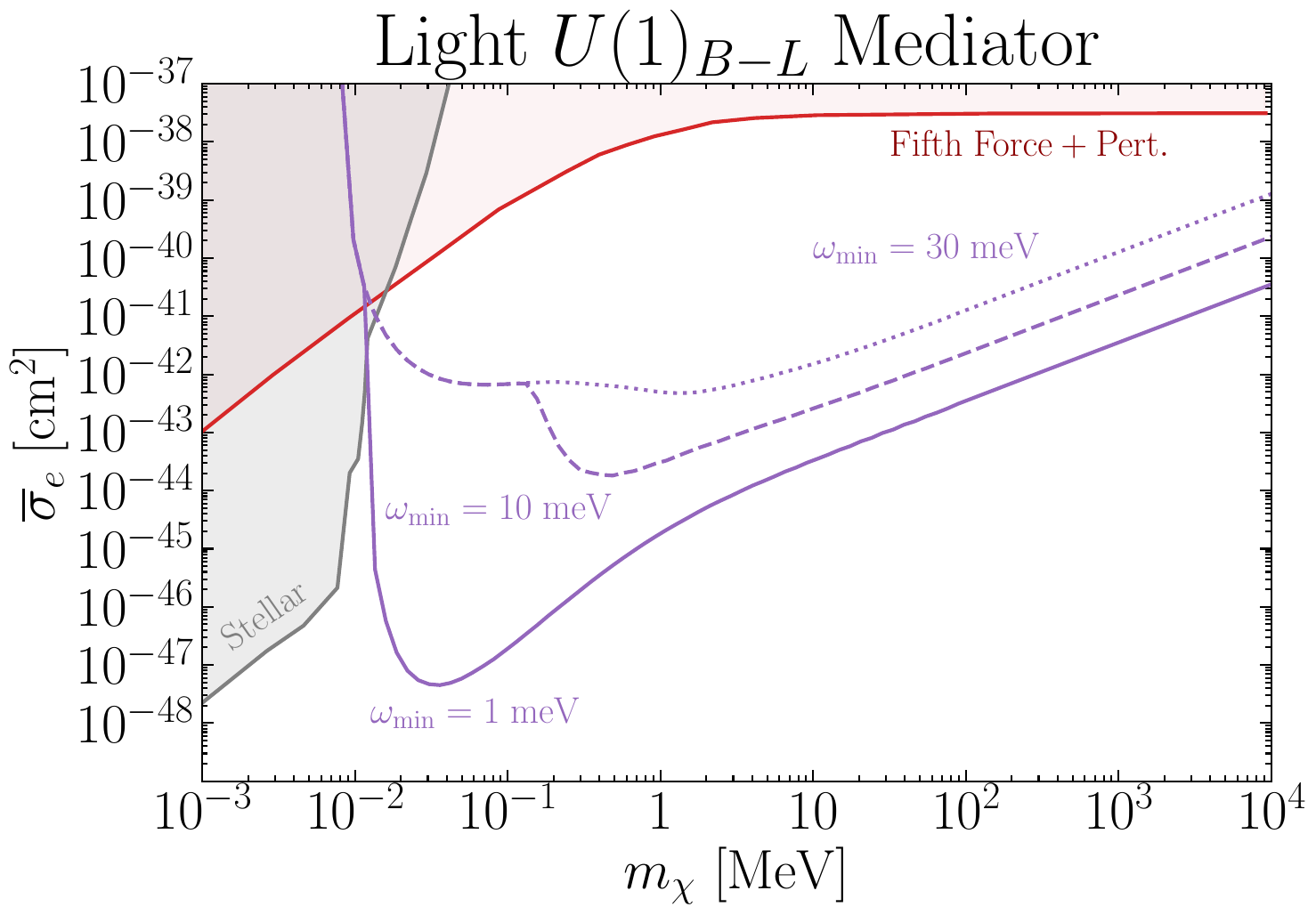}
\caption{\label{fig:sigmae}
    Projected reach for a 5\% subcomponent of DM scattering via a light (1\,eV) hadrophobic scalar (left) or $U(1)_{B-L}$ vector (right) mediator, assuming 1\,kg-yr exposure with a GaAs target, 3 signal events and no background. Single phonon excitation reach is shown in purple, assuming energy thresholds $\omega_\text{min}=1,10,30\,\text{meV}$. Pink regions are excluded when taking into account the strongest constraint on the mediator-SM coupling -- red giant (RG) stars and fifth force experiments for the two models respectively~\cite{Knapen:2017xzo} -- together with perturbativity (Pert.)\ of the mediator-$\chi$ coupling. In the $U(1)_{B-L}$ case, the gray region is excluded by stellar production of $\chi$~\cite{Vogel:2013raa}.\footref{foot:two}
}
\end{figure}

Beyond the previously studied dark photon mediator case, our first-principle rate derivation here allows us to compute the reach for other DM models with couplings to electrons. As examples, we consider two benchmark models from Ref.~\cite{Knapen:2017xzo} -- a hadrophobic light scalar mediator and a light $U(1)_{B-L}$ vector mediator. In both cases, astrophysical constraints already rule out all of the parameter space within reach of proposed experiments if $\chi$ composes all the DM. We find similar results here: for a hadrophobic light scalar mediator, the astrophysical constraints extend past the reach of single phonon excitations in a GaAs target; for a light $U(1)_{B-L}$ vector mediator, for $m_\chi \gtrsim 100$ MeV and $\omega_\text{min} = 1$ meV, the reach extends slightly past the astrophysical constraints, but the rest of the parameter space is constrained. Therefore, as in Ref.~\cite{Knapen:2017xzo}, we consider the case where $\chi$ is a 5\% subcomponent of DM, in which case SIDM constraints are absent and single phonon excitations can probe currently unconstrained parameter space. The projected reach for both benchmark models is shown in Fig.~\ref{fig:sigmae}, where a mediator mass of 1\,eV is assumed for definiteness.\footnote{\label{foot:two}Changes in the single phonon excitation reach in the left panel of Fig.~\ref{fig:sigmae}, relative to previous versions, are due to the inclusion of screening effects in DM models coupling to the electron number density via a scalar mediator. The calculations have been updated using \textsf{PhonoDark} v1.1.0~\cite{phonodark}.}

\section{Conclusions}
\label{sec:conclusions}

Dark matter direct detection has entered an era in which not only the mass coverage is extending beyond the classic WIMP window -- especially into the sub-GeV regime -- but also multi-channel target response is becoming an important consideration when designing new experiments. In this paper, we detailed a theoretical framework for calculating spin-independent direct detection rates that can be applied across multiple search channels. Starting from generic DM couplings to the proton, neutron and electron, we factored out material and channel dependent target response into the dynamic structure factor, and derived a procedure to compute this factor which involves quantizing number density operators in the appropriate Hilbert space. We focused on $\OO(\text{eV})$-gap crystal targets where existing and proposed search channels include nuclear recoils, electron transitions and single phonon excitations, each probing a different kinematic regime (see Fig.~\ref{fig:kin}). Despite the apparently very different physics involved, the calculation proceeds analogously for all three channels.

While part of this paper has been devoted to rederiving known results in this unified framework, we also obtained several new results, which we summarize in the following:
\begin{itemize}
\item We have clarified the range of validity of the standard nuclear recoils calculation (Sec.~\ref{sec:nuclear-validity}). For energy depositions lower than $\OO(100\,\text{meV})$ in a crystal target, the picture of scattering off single nuclei breaks down. Collective motions of all nuclei have to be considered, with phonons being the appropriate degrees of freedom. The situation is analogous in fluids, though the energy cutoff can be lower (e.g.\ $\OO(\text{meV})$ for superfluid helium).
\item We have extended the electron transition calculation to account for anisotropic target response, and pointed out the resulting daily modulation can be significant (Sec.~\ref{sec:electron-modulation}). As an example, we considered hexagonal boron nitride, a semiconductor with a 6\,eV gap and layered crystal structure, and showed that $\pm (10\,\text{-}\,40)\%$ daily modulation can be expected, depending on the DM mass (Fig.~\ref{fig:mod}).
\item As a major new result, we have presented a first-principle derivation of single phonon excitation rates for generic SI couplings. The final result is Eq.~\eqref{eq:Rphonon}, where dependence on the relative couplings to the proton, neutron, and electron is fully captured by the quantities $\vect{Y}_j$. Computing $\vect{Y}_j$ is straightforward for DM coupling only to nucleons (Sec.~\ref{sec:phonon-n}), but nontrivial in the presence of coupling to electrons (Sec.~\ref{sec:phonon-e}). In the latter case, we have shown that $\vect{Y}_j$ are related to the Born effective charges in an ionic crystal for a general light mediator (not necessarily a dark photon) -- see Eq.~\eqref{eq:Yj-lowq}. As examples, we computed the reach for DM scattering via a light hadrophobic scalar or $U(1)_{B-L}$ vector mediator (Fig.~\ref{fig:sigmae}), where single phonon excitations offer a complementary search channel with competitive sensitivities to previous proposals~\cite{Knapen:2017xzo}.
\item We have pointed out that sensitivity of the single phonon excitation channel is not restricted to sub-MeV DM. For heavier DM, Umklapp contribution can be significant (Fig.~\ref{fig:umklapp}), and single phonon excitations and nuclear recoils play complementary roles in probing the DM parameter space (Fig.~\ref{fig:sigman}).
\end{itemize}

In addition to shedding light on the connection and complementarity between various existing and proposed direct detection channels, the theoretical framework presented here also makes clear that there is a common algorithm one can follow to study yet unexplored novel detection channels in the future. Some of them will require extending our present formalism beyond SI interactions, a task we plan to take on in future work.

\vspace{10pt}
{\em Acknowledgments.}
We thank Thomas Harrelson, Simon Knapen, and Matt Pyle for useful discussion. T.T.\ and K.Z.\ are supported by the Quantum Information Science Enabled Discovery (QuantISED) for High Energy Physics (KA2401032) at LBNL. Z.Z.\ is supported by the NSF Grant PHY-1638509 and DoE Contract
DE-AC02-05CH11231. Computational resources were provided by the National Energy Research Scientific Computing Center and the Molecular Foundry, DoE Office of Science User Facilities supported by the Office of Science of the U.S.\ Department of Energy under Contract No.~DE-AC02-05CH11231. The work performed at the Molecular Foundry was supported by the Office of Science, Office of Basic Energy Sciences, of the U.S. Department of Energy under the same contract number. T.T.\ and Z.Z.\ would like to thank the Walter Burke Institute for Theoretical Physics for hospitality during the completion of this work.

\appendix

\section{DFT Calculation Details for BN}
\label{app:dft}

We used the Vienna \textit{Ab initio} Simulation Package (VASP)~\cite{Kresse1993,Kresse1994,Kresse1996,Kresse1996a} for our density functional theory calculations to obtain the electronic properties of BN. Projector augmented wave (PAW) pseudopotentials~\cite{Blo,Kresse1999} with the Perdew-Becke-Ernzerhof (PBE) exchange-correlation functional~\cite{Perdew1996} were used. We included van der Waals interactions between BN layers using the D3 correction method of Grimme \textit{et al.}\ with Becke-Johnson damping~\cite{Grimme2010,Grimme2011}. In the PAW scheme, we treated \textit{s} and \textit{p} electrons as valence for both B and N. 

For structural optimization, we use an energy cutoff of 950 eV for our plane wave basis set, with a Gamma-centered k-point grid of $12\times12\times12$. The total energy and forces were converged to $1\times10^8$ eV and 1 meV/\AA\ respectively. Wavefunctions were evaluated on two Gamma-centered k-point meshes, $10\times10\times3$, and $14\times14\times4$, converging the scattering rate to $\sim 9\%$ at 5 MeV, $\sim 8\%$ at 10 MeV and $\sim 6\%$ at 100 MeV. We extracted the all-electron wavefunction coefficients from our PAW calculations using pawpyseed~\cite{2019arXiv190411572B} with an energy cutoff of 450 eV. 68 energy bands were included, incorporating energies up to 60 eV above and below the valence band maximum. 

Boron nitride (BN) adopts a hexagonal crystal structure with space group $P6_{3}/mmc$ (No. 194) as shown in Fig.~\ref{fig:bn}. Our calculated lattice parameters are $a = 2.507$ \AA\ and $c = 7.093$ \AA\, which compare well to those from experiment~\cite{Lynch1966} ($a = 2.504$ \AA\ and $c = 6.661$ \AA). The PBE-level calculated band gap is 3.61 eV which was corrected to the experimental value of 5.97 eV~\cite{Watanabe2004} using a scissors operator.

\bibliographystyle{apsrev4-1}
\bibliography{refs_th-short}
\end{document}